\documentclass{aa}

\usepackage{graphicx}
\usepackage{color}
\usepackage[varg]{txfonts}
\usepackage{epstopdf}
\newcommand{\degree}{\ensuremath{^\circ}}
\usepackage{threeparttable}

\usepackage{lscape}
\usepackage{natbib}
\graphicspath{{figures/}}

\begin{document}

%%%%%%%%%%%%%%%%%%%%%%%%%%%%%%%%%%%%%%%%%%%%%%%%%%%%%%%%%%%%%%%%%%%%%%%%%%%%%%%%
%%%%%%%%%%%%%%%%%%%%%%%%%%%%
\title{Eigenspectra of solar active region long-period oscillations}

\author{G.~Dumbadze\inst{1,2,3}, B.M.~Shergelashvili\inst{2,3,4,5}, S.~Poedts\inst{1,9}, T.V.~Zaqarashvili\inst{3,6,7}, M.~Khodachenko\inst{4,8,10}, and P.~De Causmaecker\inst{5}}

\institute{Centre for mathematical Plasma Astrophysics, Department of Mathematics, KU Leuven, Celestijnenlaan 200B, B-3001, Leuven, Belgium\\
                 \and
         Centre for Computational Helio Studies, Ilia State University, G.\ Tsereteli street 3, 0162 Tbilisi, Georgia\\
                  \and
         Evgeni Kharadze Georgian National Astrophysical Observatory, M.\ Kostava street 47/57, 0179 Tbilisi, Georgia\\
                 \and
         Space Research Institute, Austrian Academy of Sciences, Schmiedlstrasse 6, 8042 Graz, Austria\\
                  \and
         Combinatorial Optimization and Decision Support, KU Leuven campus Kortrijk, E.\ Sabbelaan 53, 8500 Kortrijk, Belgium\\
                  \and
         Institute of Physics, IGAM, University of Graz, Universit\"atsplatz 5, 8010 Graz, Austria\\
                  \and
         Ilia State University, Cholokashvili Ave 3/5, 0162 Tbilisi, Georgia\\
                  \and
         Skobeltsyn Institute of Nuclear Physics, Moscow State University, Moscow 119992, Russia\\
                  \and
        Institute of Physics, University of Maria Curie-Sk{\l}odowska, Pl.\ M.\ Curie-Sk{\l}odowska 5, PL-20-031 Lublin, Poland\\
                  \and
        Institute of Astronomy, Russian Academy of Sciences, Moscow 119017, Russia
}

%%%%%%%%%%%%%%%%%%%%%%%%%%%%%%%%%%%%%%%%%%%%%%%%%%%%%%%%%%%
\abstract
{We studied the low-frequency $\lesssim 0.5\;$h$^{-1}$ (long-period $\gtrsim 2\;$h) oscillations of active regions (ARs). The investigation is based on an analysis of a time series built from Solar Dynamics Observatory/Helioseismic and Magnetic Imager (SDO/HMI) photospheric magnetograms and comprises case studies of several types of AR structures. }
{The main goals are to investigate whether ARs can be engaged in long-period oscillations as unified oscillatory entities and, if so, to determine the spectral pattern of such oscillations. }
{Time series of characteristic parameters of the ARs, such as, the total area, total unsigned radial magnetic flux, and tilt angle, were measured and recorded using the image moment method. The power spectra were built out of Gaussian-apodised and zero-padded datasets.  }
{There are long-period oscillations ranging from 2 to 20 h, similarly to the characteristic lifetimes of super-granulation, determined from the datasets of the AR total area and radial magnetic flux, respectively. However, no periodicity in tilt angle data was found. }
{Whatever nature these oscillations have, they must be energetically supported by convective motions beneath the solar surface. The possible interpretations can be related to different types of magnetohydrodynamic (MHD) oscillations of the multi-scale structure of the AR magnetic field, which is probably linked with the characteristic turnover timescales of the super-granulation cells. The presence of oscillations in the radial magnetic flux data may be connected to periodic flux emergence or cancellation processes.}

%%%%%%%%%%%%%%%%%%%%%%%%%%%%%%%%%%%%%%%%%%%%%%%%%%%%%%%%%%%
\keywords{Sun: magnetic fields; Active Regions; Methods: data analysis; Sun: oscillations}

\titlerunning{Eigenspectra of active region long-period oscillations}

\authorrunning{Dumbadze et al.}

%%%%%%%%%%%%%%%%%%%%%%%%%%%%%%%%%%%%%%%%%%%%%%%%%%%%%%%%%%%%
\maketitle

%%%%%%%%%%%%%%%%%%%%%%%%%%%%%%%%%%%%%%%%%%%%%%%%%%%%%%%%%%%
\section{Introduction}
%%%%%%%%%%%%%%%%%%%%%%%%%%%%%%%%%%%%%%%%%%%%%%%%%%%%%%%%%%%
Active regions (ARs) are areas of strong magnetic field concentration visible on the solar surface and in its lower atmosphere. These are areas of intense and often complex magnetic activity in which sunspots frequently form. ARs are often related to solar activity; for example,\ solar flares and coronal mass ejections. The number, the locations, and the sizes of these ARs vary with time, making ARs valuable indicators or tracers of the magnetic activity of the Sun. The available high-resolution (both in time and space) observations vividly show the complexity of the AR morphology and their intriguing dynamics. One important aspect of the observable dynamical properties of ARs and of the sunspots they contain concerns the huge variety of waves and oscillations they exhibit. The various oscillatory phenomena can be classified in different types such as umbral chromospheric oscillations with a typical period of three minutes \citep{Chorley10,centeno06,fleckschmits91,kuridze09}, umbral photospheric oscillations with a typical period of five minutes \citep{thomas84,shergelashvili05}, long-period oscillations with typical periods of several hours \citep{Dumbadze17,efremov07,goldvarg05}, and ultra-long-period oscillations with typical periods of several days \citep{khutsishvili98,gopasyuk04}. The long-period oscillations in sunspots were discovered both in radio emission measurements and in magnetic field data \citep{gelfreichetal06,efremov07,solovev08,Smirnova13,Abramov2013}. The ground-based and space-based observations yielded very similar long-period oscillations of sunspots that confirm the solar nature of these oscillations \citep{Abramov2013,Nagovitsyna2011}. State-of-the-art, space-based observational facilities provide data with unprecedented accuracy, which encourages both observational and theoretical studies of the physics behind AR dynamics \citep{Moradi12}.

The image moment method represents a tool for image processing to recognise visual patterns of the image objects and to characterise their position, size, and orientation \citep{flusser07, hu62}. The invariants are computed using their boundary shape and interior area, and provide characteristics of an object to represent its properties \citep{hu62,Prokop92}. \citet{hu62} used the mathematical foundation for two-dimensional moment invariants to derive seven known invariants with respect to the rotation of 2D objects. Furthermore, Hu's moment invariant method has been used in template matching and the registration of satellite images \citep[e.g.][]{Goshtasby1985,Flusser1994} for character recognition of objects \citep{BELKASIM1991,TSIRIKOLIAS1993} and in many other applications.

In \citet[hereafter Paper I]{Dumbadze17}, we reported and discussed long-period oscillations in ARs, with periods of up to several hours. In that paper, we also focused on oscillations of the tilt angle of the AR's cross-sectional area. The tilt angle oscillations were interpreted as the second harmonic of standing kink modes, assuming the ARs are related to magnetic flux tubes with feet embedded in the dense plasma of the solar interior and connected to the lighter (less dense) part of the loop observed above the surface. We assumed that the related kink modes have nodes in the apex above the solar surface and that the ARs oscillate as a whole. This simplifying assumption enabled us to roughly determine the possible distribution of their phase speed values along the corresponding magnetic tubes. Our estimations showed that the characteristic depth of the ARs is about $40\,000\;$km, and that below this depth the sunspots are presumably fragmented into smaller magnetic tube structures as was suggested by \citet{parker1979}.

The analysis presented here is in some sense a continuation of the previous phenomenological study of  ARs reported in Paper I. The objective of the present paper is to more systematically examine the existence of the oscillations in  different types of ARs. For this purpose, we performed a set of case studies for typical ARs along their transit across the solar disk, which were selected according to their morphological structure. The remainder of this paper is organised as follows: the observations of the ARs and the data processing methods are described in Sect.~\ref{secobsdata}. Section~\ref{comparison} is dedicated to the analysis of the significant periods discovered. The discussion and conclusions are presented in Sect.~\ref{secresults}.

%%%%%%%%%%%%%%%%%%%%%%%%%%%%%%%%%%%%%%%%%%%%%%%%%%%%%%%%%%%
\section{Observations and data processing methodology}\label{secobsdata}
%%%%%%%%%%%%%%%%%%%%%%%%%%%%%%%%%%%%%%%%%%%%%%%%%%%%%%%%%%%
\subsection{Observational data}\label{secdata}
%%%%%%%%%%%%%%%%%%%%%%%%%%%%%%%%%%%%%%%%%%%%%%%%%%%%%%%%%%%

\begin{figure}[!ht]
\centering
\includegraphics[scale=0.43,trim=0.83 2.93 9.57 6, clip]{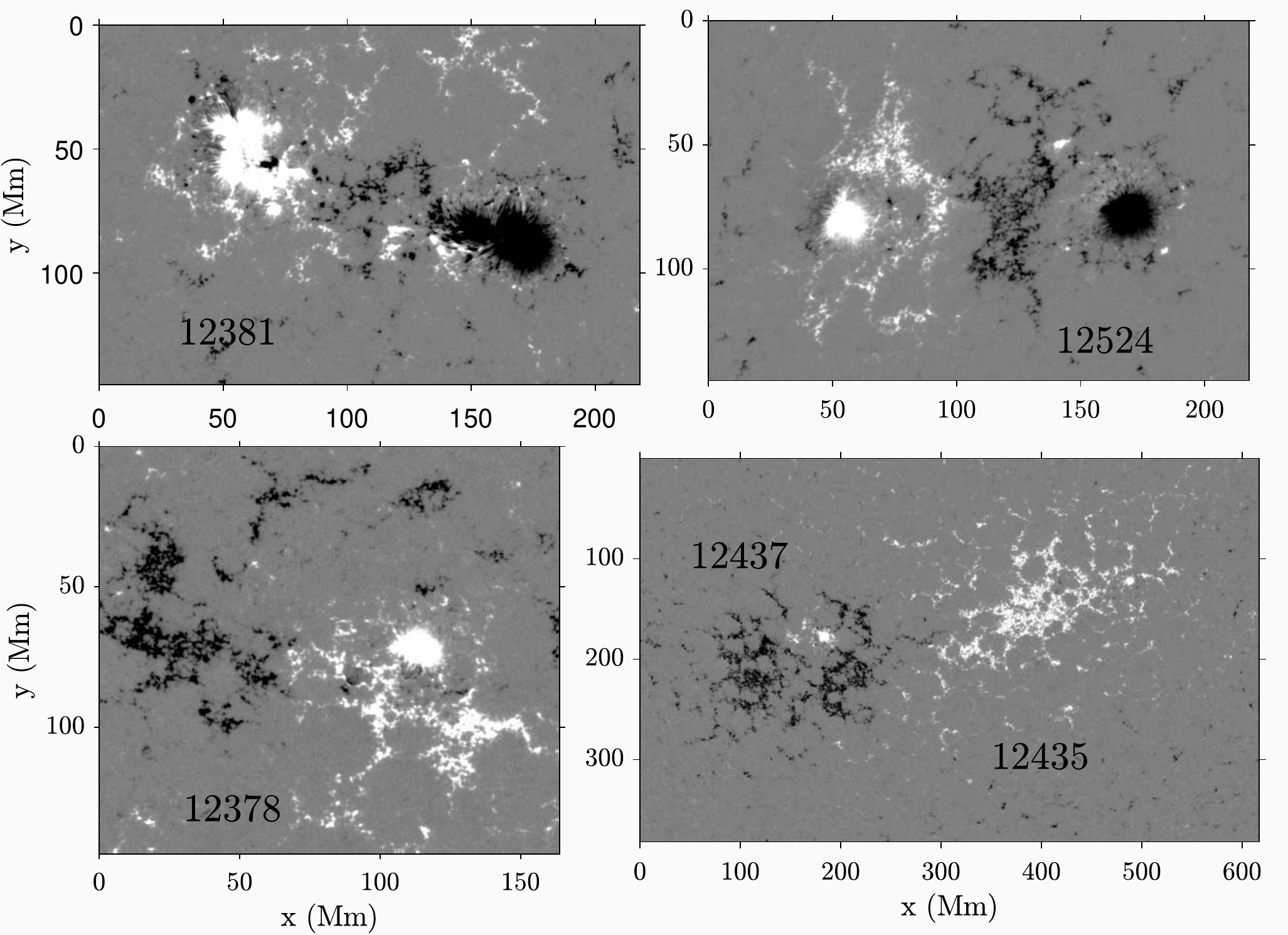}
\caption{Illustration of Helioseismic and Magnetic Imager (HMI)  magnetogram snapshots of the five studied ARs. }
\label{ARexample}
\end{figure}

The aim of the study reported here is to reveal temporal oscillatory patterns in the datasets related to a number of ARs. We carried out case studies for the five ARs that are shown in Fig.~\ref{ARexample}. The list of these five studied ARs with some of their properties is also given in Table~\ref{table1}. During the selection of these specific ARs, we paid attention to their sizes, shapes, and locations on the solar disk. The observational time span is chosen so that we observe the transition of the ARs from roughly $-70\degree$ to $+70\degree$ longitude.

\begin{table}[!ht]
\caption{The selected ARs. }
 \centering
 \scriptsize
 \begin{threeparttable}
\begin{tabular}{cccrrc}
 \hline \hline
1 warning

AR & Obs. time & \multicolumn{1}{c}{Longitude $^{(1)}$}    & \multicolumn{1}{c}{Latitude $^{(1)}$}    & \multicolumn{1}{c}{Area $^{(2)}$}     & Threshold $^{(3)}$ \\
                no.   & (h)   & \multicolumn{1}{c}{($\degree$)}  & \multicolumn{1}{c}{($\degree$)} & \multicolumn{1}{c}{(km$^2$)} & (G)   \\
 \hline
12381 & $240$ & [$-69.9\;$  $65.2$] & [$14.7\;$   $14.5$]  & $1.6\cdot10^9$ & $400$ \\
12378 & $240$ & [$-64.5\;$  $60.2$] & [$-15.7\;$  $-14.9$] & $1.5\cdot10^8$ & $300$ \\
12435 & $225$ & [$-62.5\;$  $54.4$] & [$-15.3\;$  $-15.4$] & $6.0\cdot10^7$ & $350$ \\
12437 & $225$ & [$-57.9\;$  $59.2$] & [$-21.6\;$  $-21.7$] & $6.0\cdot10^7$ & $350$ \\
12524 & $240$ & [$-69.2\;$  $69.6$] & [$14.9\;$   $14.3$]  & $6.1\cdot10^8$ & $400$ \\
 \hline
\end{tabular} \label{table1}
    \begin{tablenotes}
\item \textbf{Notes.} $^{(1)}$The location of the AR on the solar surface at the beginning and the end moment of the observation period. $^{(2)}$The area values are taken at the moment when the centre of mass of the AR is at or very close to the visible central meridian. $^{(3)}$The thresholds of unsigned magnetograms for the corresponding unsigned binary images are used to isolate magnetically active pixels \citep{Hoeksema14} for each AR.
    \end{tablenotes}
  \end{threeparttable}
\end{table}

We used data from the Solar Dynamics Observatory (SDO)/Helioseismic and Magnetic Imager (HMI) \citep{Schou12, Scherrer12}. The line-of-sight magnetogram ($B_{LOS}$) dataset of 45 s cadence  was used for the systematic investigation of active region long-period oscillatory patterns. The projection effects were corrected using the 'mtrack' algorithm in the SDO/HMI ring-diagram pipeline \citep{Bogart2011a, Bogart2011b}, which employs the azimuthal equidistant (Postel) projection. This involves a mapped rectangular frame following the AR at selected heliographic coordinates and tracks it during the time interval with the Carrington rotation rate.
Also, we considered all three components of the magnetic field (i.e.\ radial $B_r$, meridional $B_{\theta,}$ and azimuthal $B_{\varphi}$), which were obtained from Space-weather HMI Active Region Patches (SHARP, \citet{Bobra14}), available as sharp\_cea\_720s data with a cadence rate of 720 s. For these data, the projection effects were corrected using cylindrical equal-area mapping.

%%%%%%%%%%%%%%%%%%%%%%%%%%%%%%%%%%%%%%%%%%%%%%%%%%%%%%%%%%
\subsection{Generation of time series}\label{secseries}
%%%%%%%%%%%%%%%%%%%%%%%%%%%%%%%%%%%%%%%%%%%%%%%%%%%%%%%%
As mentioned before, we used the image moment method to process the obtained series of  rectangular frames described in the previous sub-section. The standard procedure implies a transformation of the original $B_{LOS}$ magnetograms (or $B_r$, $B_{\theta}$ and $B_{\varphi}$) into unsigned binary images of the ARs using certain thresholds for each of them (cf.\ last column of Table~\ref{table1}). For each of the different ARs, the value of the threshold is chosen individually from the $\pm~300-400\;$G range (see Table~\ref{table1}) so that the spurious small magnetic features around the AR are filtered out, while the principal components of the AR body remain visible, and  more or less unchanged throughout the observational time span. These thresholds are kept the same within the duration of observation for a given AR. After the implementation of thresholding and binary imaging procedures, the binary snapshots were used for the calculation of the image moments. The moments of invariants are characterised with their two-dimensional $(p+q)$th order defined by \citet{hu62}:
\begin{equation}\label{moment}
    M_{pq}=\iint{x^py^qI(x,y)\;dx\,dy},
\end{equation}
where $p,q=0,1,2,...$; $x$ and $y$ are the coordinates of the pixel in the respective scaling; $I(x,y)$ denotes the binary rate (set of active pixels) of the unsigned magnetic field. For clarity with regard to the analysis, we introduce the notation for total areas as $D_{i}$ ($i=$ $B_{LOS}$, $B_r$, $B_{\theta}$ or $B_{\varphi}$), which are defined by the corresponding unsigned magnetic field  binary rate $I$.

For the analysis, we use the discretised counterpart of Eq.~(\ref{moment}):
\begin{equation}\label{sum}
    M_{pq}=\sum_{\hbox{\tiny active pixels}}{{x^py^qI(x,y)}},
\end{equation}
where $I(x,y)$ represents the mentioned discrete binary rate of $B_{LOS}$(or $B_r$, $B_{\theta}$, $B_{\varphi}$). In this case, the zeroth-order moment invariant $M_{00}$ determines the area of the object measured in the pixel areas. We can also calculate the principal axes of the object \citep{hu62} using the first- and second-order moment invariants. Here, the principal axes of the object correspond to the major and minor axes of the AR. The inclination of the AR major axis with respect to the equatorial line of the solar disk, which is identified as the tilt angle of the major axis (the largest principal axis of the object), can be derived using the first- and second-order moments of the image object:
\begin{equation}\label{tilt}
\theta=\frac{1}{2}\arctan\left(\frac{2\left(M_{11}/M_{00}-\bar{x}\cdot \bar{y}\right)}{\left(M_{20}-M_{02}\right)/M_{00}-\left(\bar{x}^2-\bar{y}^2\right)}\right).
\end{equation}
Using the sequence of snapshots within the observational time span, we thus produced time series for the tilt angles and the total areas of the studied ARs, and this for all components of the magnetic field.
In addition, we produced time series of the total unsigned radial magnetic flux, determined as
\begin{equation}
|\Phi|=\sum_{\hbox{\tiny active pixels}} |B_{r}(x,y)|S(x,y),
\end{equation}
where $B_{r}(x,y)$ is the radial magnetic field component in each active pixel identified by the corresponding binary image, and $S(x,y)=1.33\times10^5\;$km$^2$ is the area covered by these pixels on the solar surface corresponding to the cylindrical equal area remapping \citep{Bobra14}.

%%%%%%%%%%%%%%%%%%%%%%%%%%%%%%%%%%%%%%%%%%%%%%%%%%%%%%%%
\subsection{Data pre-processing }\label{methods}
%%%%%%%%%%%%%%%%%%%%%%%%%%%%%%%%%%%%%%%%%%%%%%%%%%%%%%%%
We focussed our study on long-period ($>2\;$h) (low-frequency) oscillation signals with a satisfactory level of confidence. Further discussion reveals that such periodicities are seen in the Fourier spectra of the considered time series. However, they are 'drenched' in the background power-law noise, making it difficult to prove the significance of the detected periods. Therefore, in order to elaborate the characteristic periods with a satisfactory level of confidence and to distinguish the real quasi-oscillations in the time series from artefacts, we used three different methods for the period detection:

\subsubsection{Method 1 (M1) -- Power spectra of the datasets}
\citet{Vaughan2005} suggested a method that finds significant peaks on top of the power-law noise and calculates their uncertainties. We applied this method to our datasets to find significant peaks in the power spectra, implying the following steps. (i)~De-trending and apodisating the initial data (as described in Appendix~\ref{appmethod}, items 1\ and 2). (ii)~Computing of the periodograms of the detrended and apodised datasets; and along with this, we omit the power of the Nyquist frequency following the recipe given in \citet{Vaughan2005} and convert the frequency ($f$) and power ($W$) into log-space (see example in Fig.~\ref{M1}). (iii)~Making a linear model fit ($\lg W_{model}=m\lg f+b$) of the obtained power spectrum in log-scale (cf.\ solid blue lines in Fig.~\ref{M1}). (iv)~Determining the residuals of the computed power from the model line assuming that they follow the second order two-dimensional chi-square ($\chi_2^2$) distribution as a null hypothesis. In particular, we test the goodness of the model with the Kolmogorov-Smirnov (KS) test \citep{press2007}. When the KS test does not satisfy the null hypothesis, we adjust the slope $m$ and the offset $b$ until the test is fulfilled. (v)~Estimating the 95\% probability limit $\gamma_e$ by assuming that the noise is two-dimensional $\chi^2$ distributed following Eq.~(16) from \citet{Vaughan2005} and using the computed $m$ and $b$ coefficients (we also plot this confidence limit into log-space; cf.\ dashed red lines in Fig.~\ref{M1}). (vi)~Selecting the significant peak of the lowest frequency (longest period) in the spectrum and comparing it with similar period of the sinusoidal fitting function $A_1\sin(A_2t+A_3)$ applied to the considered initial dataset (for further details, we refer the reader to Appendix~\ref{appmethod}). The corresponding peak is then removed from the power spectrum by directly subtracting the found model sine function from the dataset. (vii)~Steps (ii)-(v) are iteratively repeated until all outstanding peaks are removed from the spectrum, and a pure power-law noise remains.\\[6pt]

\subsubsection{Method 2 (M2) -- Re-binned power spectra of the datasets}
In order to enhance significant peaks in the power spectrum, we used the spectral re-binning method \citep{Appourchaux2003}. \citet{Pugh2017} showed how the \citet{Vaughan2005} method can be updated for use with re-binned power spectra. They used this  method for solar and stellar flare datasets to determine quasi-periodic pulsations. We used the same method and proceeded with the following steps: (i)~performing steps (i) and (ii) of M1; (ii)~compute the sum of consecutive $p$ frequency bins using the original power spectrum and then divide it by $p$ (in our calculations we use $p=2$); (iii)~repeat steps (iii)-(iv) of M1 (however, for the description of the noise distribution, instead of a two-dimensional $\chi_2^2$ distribution function, we used a $2p$-dimensional distribution function $\chi_{2p}^2$  \citep{Pugh2017}; then, the corresponding 95\% confidence line is estimated and plotted accordingly); (iv)~repeat steps (vi)-(vii) of M1 (see example in Fig.~\ref{M1}).\\[6pt]

\subsubsection{Method 3 (M3) -- Average power spectra for four consecutive equal parts (time windows) of entire observational time}
In order to provide additional proof of the presence of the discovered significant periods in the signal, we used a third method. It is based on the division of the total observational time span in four equivalent mutually non-overlapping intervals of time, and we derived the power spectra for each of these intervals. Considering the shorter datasets, we sacrifice the spectral resolution (the uncertainties increase by a factor four compared to the first two methods) and, in this case, the noise no longer follows a chi-squared distribution. However, we are able to generate the 'averaged' Fourier power spectrum out of these four observational windows.

The analysis steps of M3 are as follows. (i)~Divide the analysed time series into four equivalent non-overlapping parts. (ii)~Compute the power spectrum of each individual part, as described in step (ii) of M1. (iii)~Sum the four power spectra obtained. The peaks revealed by M1 and M2 are 'real' if they  still remain in the summed spectrum, while the spurious ones disappear (see example in Fig.~\ref{M1}). Then, we apply the same procedures as those outlined in steps (iii)-(vii) of M1 again.

Both methods M1 and M2 enable us to isolate the individual periods of oscillations with a spectral accuracy of approximately the same magnitude. The purpose of using these two different methods is to cross-check the validity of the discovered significant peaks. Thus, we are able to compare the periods obtained from these two main methods and evaluate their average values. Strictly speaking, however, these two methods of measurement are not completely independent of each other. Therefore, when we calculate the mean values of the periods, including measurements of areas and fluxes related to particular components of the magnetic field obtained by one of the two methods and cross confirmed by them where possible, the aim is not to demonstrate the statistical mean of the mutually independent ensemble of the measurements (according to its classical definition). Instead, we use the average values just as a method of approximation to evaluate the globally averaged values we discovered. With regard to the errors, the simple averaging of uncertainties, when we take the arithmetic mean of measurements, is not consistent. In particular, it is not consistent in the case when the methods of measurements are not completely independent of each other. Therefore, we choose the uncertainty approximation method in order to be prudent with the assumption that the uncertainty of the mean of two measurements cannot be larger than the maximum error of each individual method; hence, whenever we use the maximum uncertainties. Additionally, M3 enables the verification and confirmation (with a reduced but still reasonable accuracy) of the characteristic periods inferred by the main methods, M1 and M2, where possible. This yields the maximal robustness of the analysis that can be achieved.

%%%%%%%%%%%%%%%%%%%%%%%%%%%%%%%%%%%%%%%%%%%%%%%%%%%%%%%%%%%
\section{Analysis and comparison of spectra for different cases}\label{comparison}
%%%%%%%%%%%%%%%%%%%%%%%%%%%%%%%%%%%%%%%%%%%%%%%%%%%%%%%%%%%

%%%%%%%%%%%%%%%%%%%%%%%%%%%%%%%%%%%%%%%%%%%%%%%%%%%%%%%%%%%
\subsection{Revealing of major periodicities}\label{spectra}
%%%%%%%%%%%%%%%%%%%%%%%%%%%%%%%%%%%%%%%%%%%%%%

%%%%%%%%%%%%%%%%%%%%%%%%%%%%%%%%%%%%%%%%%%% Figures
\begin{figure*}[!ht]
\center{\includegraphics[width=0.98\linewidth]{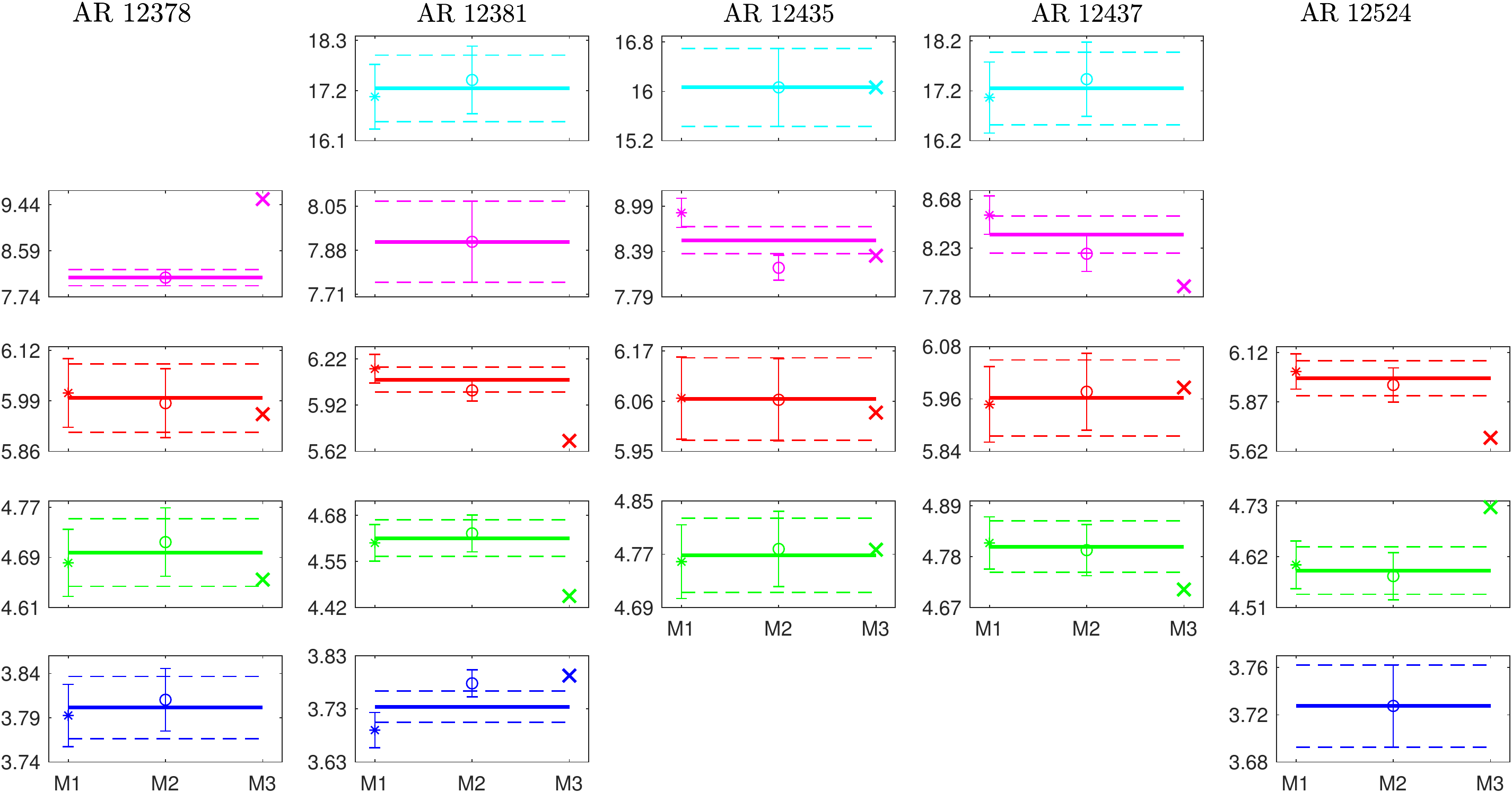}}
\caption{Characteristic mean oscillation periods (measured in hours) calculated as a global average of those found in different time series separately (for $D_{B_{LOS}}$, $D_{B_{r}}$, $D_{B_{\theta}}$, $D_{B_{\varphi,}}$ and total radial magnetic flux). On the horizontal axes, the M1, M2 and M3 labels, respectively, denote the global mean of the periods obtained by M1 (coloured '$*$' with error bars), M2 (coloured '$\circ$' with error bars), and M3 (coloured '$\times$' with error bars). Moreover, we also indicate the total mean periods by taking the average of the mentioned global means of M1 and M2 (i.e.\ the mean periods obtained by M3 are excluded from the total mean calculation) via the solid-coloured
horizontal lines (with the uncertainties indicated by the dashed coloured
lines). The colour scheme follows the sequence of quasi-harmonics described in Sect.~\ref{eigenmodes} and also corresponds to the colouring used in the tables (see Appendix~\ref{append}).}
\label{errbar1}
\end{figure*}
%%%%%%%%%%%%%%%%%%%%%%%%%%%%%%%%%%%%%%%%%%%--6
\begin{figure*}[!ht]
\center{\includegraphics[width=0.98\linewidth]{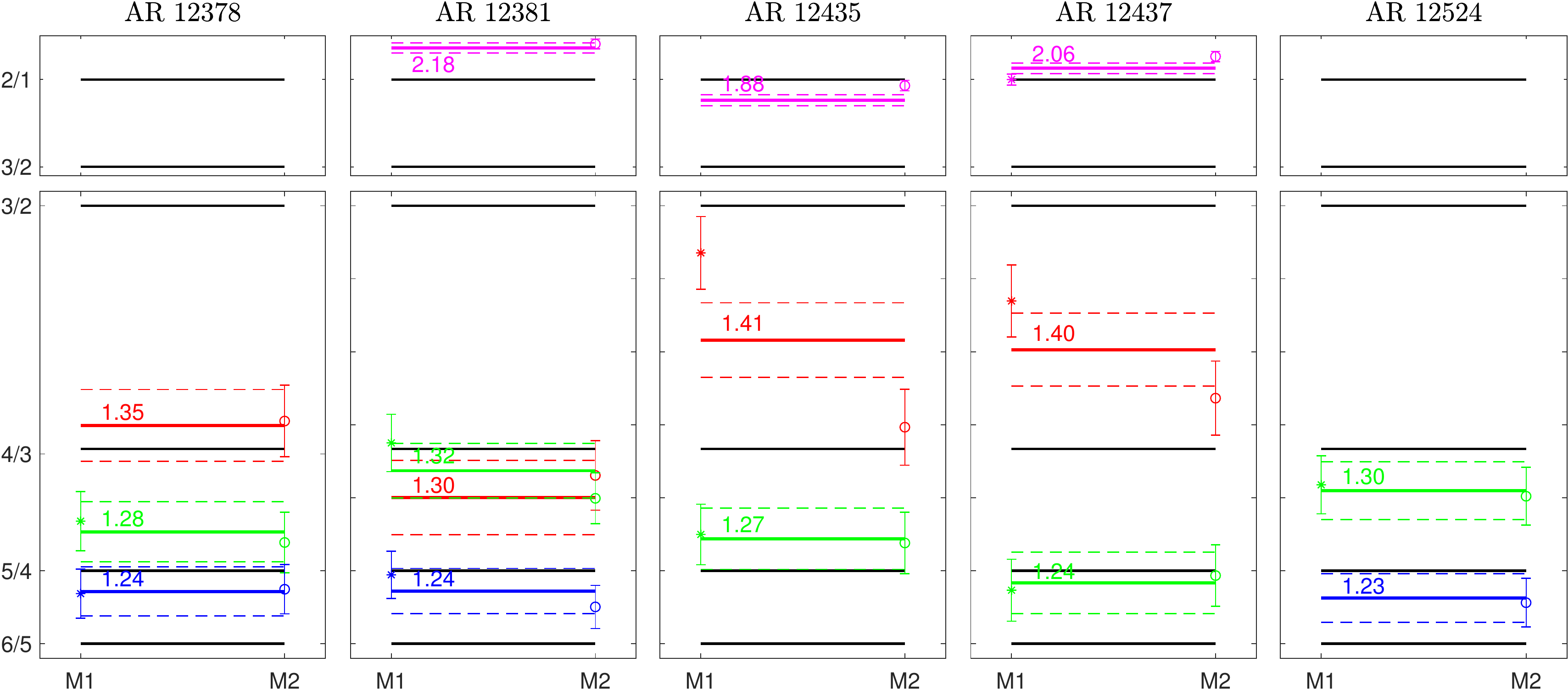}}
\caption{Ratios of global mean periods ($P_{i}/P_{i+1}$, $i=1,2,\ldots$) of M1, M2 and the total mean periods shown in the panels of Fig.~\ref{errbar1}. The formatting of the data points and horizontal lines is the same as in Fig.~\ref{errbar1}. The colouring in each panel coincides with the period standing in the denominator in each ratio. On the vertical axes, the levels of the period ratios in the reference spectrum are labelled and the corresponding horizontal solid black lines are plotted in all panels.}
\label{harmoniccs1}
\end{figure*}
%%%%%%%%%%%%%%%%%%%%%%%%%%%%%%%%%%%%%%%%%%--7

We study the long-period ($>2$~h) oscillatory patterns in ARs with four different typical morphological structures \citep[e.g.\ following the classification given in][]{Jaeggli2016}. The considered periods are obtained from the Fourier spectra of the time series, which have certain (limited) observational time spans and sampling rates. The characteristic parameters of these datasets determine the period resolution and the overall accuracy of the obtained power spectra. As a result, some significant periods in the spectra are embedded in strong background noise showing a power-law nature (see Appendix~\ref{appfigures}).
We do not attempt to provide a physical interpretation of the mechanism(s) that might be responsible for the appearance of these oscillatory patterns. Instead, we give a few plausible indications concerning the possible  scenarios. The final interpretation of the detected oscillations requires the construction of rigorous mathematical models, and this is beyond the scope of the present, purely observational, study. The results of the performed analysis for each considered AR are summarised in Tables~\ref{AR2378}-\ref{flux} (see Appendix~\ref{append}).

%%%%%%%%%%%%%%%%%%%%%%%%%%%%%%%%%%%%%%%%%%%%%%%%%%%%%%%%%%%
Most of the periodic features are revealed for $D_{B_{LOS}}$. However, the analysis of the estimated $D_{B_{r}}$, $D_{B_{\theta}}$ and $D_{B_{\varphi}}$ and the total unsigned radial magnetic flux yield a smaller number of  detected significant periods. In most cases, the values of these periods are similar to those in $D_{B_{LOS}}$.

Tables~\ref{AR2378}-\ref{flux} show that each AR yields a different number of modes, and further investigations with more ARs are needed to verify the dependence of the number of oscillation modes on the morphology. Another finding is that long periods ($>10$~h) are detected less frequently than the shorter periods ($<10$~h) in all analysed datasets. The apparent reason for this is the gradual reduction of the spectral resolution with the growth of the periods. We know also that the longest period parts of the spectra are contaminated by two artificial instrumental peaks at $12$ and $24$ h \citep{Liu12}, which might cause spectral power leakage and mixture between real, long periods and artificial spectral peaks. Presumably, these artificial signals have less impact on the part of the spectrum below $10$ h. Altogether, we show that in all datasets there are long-period oscillations with characteristic periods in the range from $2$ to $20$ h (beyond $20$ h, the derived spectra are rather unreliable because of a very large uncertainty), which are similar to the characteristic lifetimes of super-granulation \citep{Rincon2018}. So, it is intuitive to connect the observed oscillation periods with the characteristic turnover times of super-granulation cells.

None of the applied analysis methods described in Sect.~\ref{methods} gave us any  significant periodicities in the AR major axis tilt angle data. Therefore, we assume that the power spectrum for this quantity does not contain significant periods and it remains of a pure power-law type. At the same time, we would like to note that this non-detection does not mean that tilt angle oscillations do not exist, because the  uncertainty tolerance of the image moment method used here, does not allow for sufficient confidence of such oscillations. The difference between this result and that discussed in Paper~I can be related to the methodology of the analysis. In particular, the method discussed in Paper~I, identifies the principal axis of the AR and its tilt on the basis of the least-squares fitting  of its elliptical shape. While with the method of image moments used in the present paper, the tilt angle $\theta$ (Eq.~\ref{tilt}) is defined in terms of zeroth-, first-, and second-order moments of each image, which probably produces more noise in the tilt angle datasets. Thus, the noise might overwhelm the oscillation amplitude itself. Of course, a more precise analysis of this difference requires a thorough cross-comparison of the two methods of measurements.

The radial unsigned magnetic flux shows some significant periodicities (see Table~\ref{flux}) for all ARs, comparably to those of the area oscillations (see Tables~\ref{AR2378}-\ref{AR2524}). For example, the oscillations with periods of about $6$ and $4$ h are the most frequently detected ones. They are confirmed, with some exceptions, by the two main methods M1 and M2, as well as being additionally proven by M3 (with some discrepancy tolerated by the error of M3). These periods are found in the radial unsigned magnetic flux oscillations in practically all ARs. The frequent presence of the six- and four-hour periods both in area and radial magnetic flux datasets may be connected with periodic flux emergence or cancellation processes, which can be another possible component of the physical scenario behind the observed dynamic picture.

%%%%%%%%%%%%%%%%%%%%%%%%%%%%%%%%%%%%%%%%%%%%%%%%%%%%%%%%%%%
\subsection{Possible interpretation as a signature of quasi-standing oscillation patterns}\label{eigenmodes}
%%%%%%%%%%%%%%%%%%%%%%%%%%%%%%%%%%%%%%%%%%%%%%
Our analysis enables the detection of a discrete spectrum of significant periods. Tables~\ref{AR2378}-\ref{flux} (in Appendix~\ref{append}) show that the particular and averaged periods measured by various methods are grouped in the following ranges: (i) 'harmonic' $P_1$ -- $17.1\pm0.71-17.4\pm0.74$ h (marked in cyan), (ii) 'harmonic' $P_2$ -- $7.66\pm0.14-9.54\pm0.90$ h (marked in magenta), (iii) 'harmonic' $P_3$ -- $5.69\pm0.08-6.37\pm0.10$ h (marked in red), (iv) 'harmonic' $P_4$ -- $4.45\pm0.19-4.88\pm0.23$ h (marked in green), and (v) 'harmonic' $P_5$ -- $3.66\pm0.03-3.88\pm0.05$ h (marked in blue). In Fig.~\ref{errbar1}, we plot all these period groups together. In Fig.~\ref{harmoniccs1}, we show the ratios of the average periods from each group in the same sequence as in Fig.~\ref{errbar1}. For the calculation of the uncertainties found in the period ratios, we use the following relation:
$$\vartriangle\frac{P_{i}}{P_{i+1}}= \left| \frac{1}{P_{i+1}}\vartriangle P_{i}-\frac{P_{i}}{P_{i+1}^2}\vartriangle P_{i+1} \right| \leq \left| \frac{\vartriangle P_{i}}{P_{i+1}} \right| + \left| -\frac{P_{i}}{P_{i+1}^2}\vartriangle P_{i+1} \right| ,$$
where $P_{i}>P_{i+1}$ ($i=1,2,\ldots$ is the number of 'harmonics') are the global and total mean periods (see definition of these terms in the caption of Fig.~\ref{errbar1}) evaluated for each AR. They sequentially follow each other in descending order.
One can see that the ratios of the averaged periods resemble the sequence of oscillation harmonics (Fig.~\ref{harmoniccs1}) typical for a standing oscillations. The latter are shown with black solid horizontal lines as the reference values of the ratios. We note that the observed ratios do not exactly coincide with the sequence of ratios of periods corresponding to pure standing oscillations ($2/1,3/2,4/3,\ldots$). Nevertheless, the observed ratios with the entire uncertainty interval are well separated and approximately follow the reference spectrum ratios. Hence, with the present level of accuracy we can say that the observed discrete spectrum follows the sequence of quasi-standing oscillations still showing some shift of the periods and corresponding ratios from the reference values. This shift of periods might be due to the incomplete line-tying of the magnetic loop system, Doppler shifts due to the internal flows in ARs, and/or the inhomogeneity of the magnetic loop system constituting the AR below and above the solar surface. The presence of the structures in the emerged magnetic fields with power-law-distributed spatial scales is well known and actively used in sand-pile models of flares (see \citet{Baiesi2008} and references therein). The formation of such transfer structures can be ascribed to non-equilibrium statistical process of driven diffusion \citep{Maes2009} of the emerged elements of the magnetic flux.

%%%%%%%%%%%%%%%%%%%%%%%%%%%%%%%%%%%%%%%%%%%%%%%%%%%%%%%%%%%
\section{Discussion and conclusions}\label{secresults}
%%%%%%%%%%%%%%%%%%%%%%%%%%%%%%%%%%%%%%%%%%%%%%%%%%%%%%%%%%%
One can link the discovered periods with the characteristic turnover timescale of the super-granulation cells, which is known to be of the same order of magnitude. Another possible interpretation can be related to the different types of MHD oscillations of the multi-scale structure of the AR magnetic field. It is well known that the MHD oscillations and waves sustained in structures with non-uniform magnetic field can be engaged in the processes of resonant \citep{Terradas2008} or swing \citep{Shergelashvili2005} absorption. It is evident that, whatever nature these oscillations have, they must be energetically supported by convective motions beneath the solar surface, which represents the source of entropy production in the system through macroscopic stochastic super-granular motions of the plasma. The entropy variations can be due to microscopic transport processes as well \citep{Shergelashvili2007}. The presence of oscillations in the radial magnetic flux data may be connected with a periodic flux emergence/cancellation process. One could interpret some parts of the discovered spectra in terms of sequences of quasi-standing eigen-oscillations with a full or partial line-tying configuration of the magnetic structures. The MHD oscillations can be also driven by convective turbulence. To confirm which of these scenarios prevails, further dedicated studies are needed.

In the very simple set-up we considered in Paper I, we interpreted the oscillations of the major and minor axis lengths of the ellipses we used to frame the ARs as higher (than sausage mode) degree flute modes. However, this interpretation was rather intuitive and speculative. The goal of the current work is to reveal the basic spectral characteristic of the oscillations that will be used in the next work for a more rigorous modelling. The comparison of the theoretical model with the observed spectra discussed here will enable us to draw definite conclusions about the nature of these oscillations. According to our preliminary results in Paper I, one can intuitively expect that the large period oscillations should be related to the modes with phase speeds comparable to the Alfv\'{e}n velocity in the high-beta plasma in sunspots. Another general conclusion can be that, whatever nature these modes have, if for some reason there is a line-tying at some depth enabling the presence of nodes for velocity or pressure perturbations, then such discrete spectra with fixed sequences of period ratios can be formed. Since we plan further theoretical investigations of the issue, we avoided any speculations about the physical nature of the observed periodicities in the present paper, with an ultimate goal of simply addressing the observational facts; that is to say,\ the oscillation spectra  observed in the range of $2-20$~h periods. Concerning the pure standing oscillations, these would occur when the ratios of the periods of all overtones strictly following the sequence $2/1,3/2,4/3,\ldots$. The reason why the ratios differ form the reference spectrum requires investigation. Most likely, the inhomogeneity along the flux tubes causes this effect, as it does in coronal loops \citep{2009SSRv..149....3A}. Nevertheless, our results indicate that the ratios of the detected periods roughly follow the reference spectrum (shown on the ordinate axis of Fig.~\ref{harmoniccs1}); however, they still deviate from it somewhat. This might be caused by the uncertainty tolerance we admit in our measurements and the foot point leakage of modes (not complete line-tying) or their combination.

%%%%%%%%%%%%%%%%%%%%%%%%%%%%%%%%%%%%%%%%%%%%%%%%%%%%%%%%%%%
\begin{acknowledgements}
The work was supported by Shota Rustaveli National Science Foundation grants DI-2016-52, FR17\_609.  Work of G.D. was supported under Shota Rustaveli National Science Foundation grants for doctoral students -- PhDF2016\_177 and grant for young scientists for scientific research internships abroad IG/50/1/16.
Work of B.M.S. was supported by the Austrian Fonds zur Forderung der wissenschaftlichen Forschung (FWF) under the project P25640-N27. S.P. was supported by the projects C14/19/089  (C1 project Internal Funds KU Leuven), G.0D07.19N  (FWO-Vlaanderen), SIDC Data Exploitation (ESA Prodex-12).  The work of TVZ was supported by the Austrian Science Fund (FWF, project P30695-N27).
M.L.K.\ additionally acknowledges the projects I2939-N27 and S11606-N16 of the Austrian Science Fund (FWF), grant No.075-15-2020-780 (GA No.13.1902.21.0039) of the Russian Ministry of Education and Science and the project "Study of stars with exoplanets", supported by grant No.075-15-2019-1875 from the government of Russian Federation. We are thankful to anonymous referee for constructive remarks on our manuscript that led to the significant improvement of content.
\end{acknowledgements}
%%%%%%%%%%%%%%%%%%%%%%%%%%%%%%%%%%%%%%%%%%

%%%%%%%%%%%%%%%%%%%%%%%%%%%%%%%%%%%%%%%%%%
\bibliographystyle{aa}
\bibliography{mybib}

%%%%%%%%%%%%%%%%%%%%%%%%%%%%%%%%%%%%%%%%%%%%%%%%%%%%%%%%%%%
\appendix
%%%%%%%%%%%%%%%%%%%%%%%%%%%%%%%%%%%%%%%%%%%%%%%%%%%%%%%%%%%
\section{Data processing details}\label{appmethod}
%%%%%%%%%%%%%%%%%%%%%%%%%%%%%%%%%%%%%%%%%%%%%%%%%%%%%%%%%%
For the application of original, re-binned and windowed power spectra to the analysed datasets, we use data de-trending, apodisation, significant peak identification, and removal by sine function fitting, which are specified below.

%---------------------------------------------------------
\subsection{Data de-trending and preparation for different spectral methods}
%---------------------------------------------------------

Data processing begins with our obtaining  time series using the method described in Sect.~\ref{secseries} of image moments dealing with the selected values of magnetic field thresholds for individual ARs. However, apart from the stochastic and quasi-periodic parts, the initial data include aperiodic trends with a characteristic timescale of variation equal to the total observation time. We remove these aperiodic variations by subtracting a third-order polynomial function fitted to the original data. Furthermore, as mentioned in Sect.~\ref{secdata}, for 45-second cadence data of $B_{LOS}$ we use Postel's equidistant projection mapping. Additionally, we used 720-second cadence data from the Joint Science Operations Center (JSOC, \citet{jsoc2013,Hoeksema14}). The latter data are de-projected by using cylindrical equal area mapping for radial $B_r$, meridional $B_{\theta}$ and azimuthal $B_{\varphi}$ components of the magnetic field. Therefore, the projection effect related to the spherical shape of the Sun is removed from the magnetogram image data.

%---------------------------------------------------------
\subsection{Data apodisation}
%---------------------------------------------------------

We perform a Gaussian apodisation of the de-trended data to reduce the effect of finiteness of the total sampling time. The procedure implies multiplication of the original datasets by a Gaussian function of the following form:
      \begin{equation}\label{gaussian}
          w(n)=e^{-\frac{1}{2}\left(\alpha \frac{2n}{(N-1)} \right)^2},
      \end{equation}
      where $N$ is total number of sample points, while the integer $n$ is varying in the range $-(N-1)/2 \leq n \leq (N-1)/2,$ and the parameter $\alpha$ controls the shape of the Gaussian. For all area datasets of ARs 12381, 12435, 12437, and 12524, for $D_{B_{LOS}}$ of AR 12378, and for the radial flux datasets of ARs 12381, 12435, and 12524, we use $\alpha = 3$. For the remaining datasets, we use $\alpha = 2$. These $\alpha$ values were chosen for each AR separately by gradually increasing it until we obtain the value for which any further increase does not show any further significant affect on the obtained spectrum.

%---------------------------------------------------------
\subsection{Significant peak identification}
%---------------------------------------------------------
To identify the significant peaks using this method, we take the following steps:
\begin{enumerate}
\item  Perform a regression of the dataset on the sum of sine functions. We choose the number of terms in this sum depending on the best regression goodness and detect the one with the longest period (with its amplitude, angular frequency, and phase).
\item  Test the relevance of the periods found, analysing an auto-correlation function of the de-trended datasets, by taking its power spectra and making sure that the period deduced directly in step (1) is indeed present in the spectrum. Then this term is subtracted from the original dataset which almost completely removes the longest oscillation period from the signal.
\item  Take the remaining dataset, after the subtraction of the largest periodic signal found, as a new dataset and apply steps (1) and (2) to identify and extract the next longest period. These iterations are repeated until the shortest period of interest is detected and extracted.
\end{enumerate}

%%%%%%%%%%%%%%%%%%%%%%%%%%%%%%%%%%%%%%%%%%%%%%%%%%%%%%%%%%%
\section{Tables}\label{append}
%%%%%%%%%%%%%%%%%%%%%%%%%%%%%%%%%%%%%%%%%%%%%%%%%%%%%%%%%%%
Here, we present the tables that summarise the significant periods obtained from the total area variation in time corresponding to AR 12378, AR 12381, AR 12435, AR 12437, and AR 12524 (see Tables~\ref{AR2378}-\ref{AR2524}). The first columns of these tables show the periods obtained by the power spectra of the datasets (M1). The second columns indicate the periods detected by the method using the re-binned power spectra of the datasets (M2). The third columns indicate the averaged periods calculated based on the results of M1 and M2, and the last columns denote the outcome of average power spectra for four consecutive equal parts of entire observational time (M3). All periods and uncertainties shown in the tables are expressed in hours. The colours correspond to the colour-coding of the period groups described in Sect.~\ref{eigenmodes}.

In addition, we provide  Table~\ref{flux}, which contains similar periodicities inferred from the total radial magnetic flux of all considered ARs. The average characteristic parameters of the oscillations, derived from the content of these tables, are visualised in Fig.~\ref{errbar1}, while the ratios of the derived global and total mean periods are plotted in Fig.~\ref{harmoniccs1}. When multiple peaks occur in one particular method only, while a single isolated one is detected in other methods, we calculate the weighted mean of the frequency and define them as 'multiplets'.  Accordingly, we assume that such multiplets do not carry real physical information, but they simply represent spurious artefacts related to analysis accuracy.

%%%%%%%%%%%%%%%%%%%%%%%%%%%%%%%%%%%%%
%\begin{landscape}
\begin{table*}[!ht]
\caption{Detected periods (with corresponding errors) for the area of the ARs 12378 and 12381. }
\label{AR2378}
 \centering
     \begin{threeparttable}
\begin{tabular}{cccccccc}
 \hline \hline
\multicolumn{4}{c}{AR 12378 } & \multicolumn{4}{c}{AR 12381 }  \\
 \hline
M1 & M2 & Averaged periods & M3 & M1 & M2 & Averaged periods & M3  \\
 \hline
\multicolumn{4}{c}{$D_{B_{LOS}}$} & \multicolumn{4}{c}{$D_{B_{LOS}}$} \\
  \hline
-- & -- & -- & -- & \color{cyan}{$17.1\pm0.71$} & \color{cyan}{$17.4\pm0.74$} & \color{cyan}{$17.3\pm0.74$} & -- \\
-- & \color{magenta}{$8.92\pm0.20$}\tnote{*} & \color{magenta}{$8.92\pm0.20$} & \color{magenta}{$9.54\pm0.90$}\tnote{*} & -- & \color{magenta}{$7.91\pm0.15$}\tnote{*} & \color{magenta}{$7.91\pm0.15$} & -- \\
\color{red}{$5.94\pm0.09$} & \color{red}{$5.89\pm0.08$} & \color{red}{$5.92\pm0.09$} & \color{red}{$5.69\pm0.32$} & \color{red}{$6.37\pm0.10$}\tnote{*} & \color{red}{$6.14\pm0.09$}\tnote{*} & \color{red}{$6.26\pm0.10$} & -- \\
-- & \color{green}{$4.68\pm0.05$} & \color{green}{$4.68\pm0.05$} & -- & \color{green}{$4.60\pm0.05$} & \color{green}{$4.58\pm0.05$} & \color{green}{$4.59\pm0.05$} & \color{green}{$4.45\pm0.19$} \\
\color{blue}{$3.79\pm0.04$} & \color{blue}{$3.81\pm0.04$} & \color{blue}{$3.80\pm0.04$} & -- & \color{blue}{$3.69\pm0.03$} & \color{blue}{$3.67\pm0.03$} & \color{blue}{$3.68\pm0.03$} & \color{blue}{$3.79\pm0.14$} \\
$3.28\pm0.03$ & -- & $3.28\pm0.03$ & $3.10\pm0.10$ & -- & -- & -- & -- \\
-- & $2.82\pm0.04$ & $2.82\pm0.04$ & $2.84\pm0.08$ & -- & -- & -- & -- \\
\hline
\multicolumn{4}{c}{$D_{B_{r}}$} & \multicolumn{4}{c}{$D_{B_{r}}$} \\
  \hline
-- & \color{magenta}{$7.66\pm0.14$} & \color{magenta}{$7.66\pm0.14$} & -- & -- & -- & -- & -- \\
\color{red}{$6.11\pm0.09$}    & \color{red}{$5.89\pm0.08$}     & \color{red}{$6.00\pm0.09$}    & \color{red}{$6.02\pm0.36$} & -- & -- & -- & -- \\
\color{green}{$4.65\pm0.05$} & \color{green}{$4.68\pm0.06$} & \color{green}{$4.67\pm0.06$} & \color{green}{$4.65\pm0.21$} & -- & -- & -- & -- \\
 \hline
\multicolumn{4}{c}{$D_{B_{\theta}}$} & \multicolumn{4}{c}{$D_{B_{\theta}}$} \\
  \hline
-- & \color{magenta}{$8.14\pm0.17$}\tnote{*} & \color{magenta}{$8.14\pm0.17$} & -- & -- & -- & -- & --  \\
\color{red}{$5.77\pm0.08$} & \color{red}{$5.95\pm0.09$}\tnote{*} & \color{red}{$5.86\pm0.09$} & \color{red}{$6.02\pm0.36$} & \color{red}{$5.94\pm0.08$} & \color{red}{$5.89\pm0.08$} & \color{red}{$5.92\pm0.08$} & \color{red}{$5.69\pm0.63$} \\
\color{green}{$4.72\pm0.05$} & \color{green}{$4.78\pm0.06$}\tnote{*} & \color{green}{$4.75\pm0.06$} & \color{green}{$4.65\pm0.21$} & -- & -- & -- & --  \\
-- & -- & -- & -- & -- & \color{blue}{$3.88\pm0.05$} & \color{blue}{$3.88\pm0.05$} & --  \\
 \hline
\multicolumn{4}{c}{$D_{B_{\varphi}}$} & \multicolumn{4}{c}{$D_{B_{\varphi}}$} \\
  \hline
-- & \color{magenta}{$7.66\pm0.14$} & \color{magenta}{$7.66\pm0.14$} & -- & -- & -- & -- & -- \\
\color{red}{$6.02\pm0.09$} & \color{red}{$5.95\pm0.09$}\tnote{*} & \color{red}{$5.99\pm0.09$} & \color{red}{$6.02\pm0.36$} & -- & -- & -- & -- \\
-- & -- & -- & -- & -- & \color{green}{$4.68\pm0.03$} & \color{green}{$4.68\pm0.03$} & -- \\
 \hline
\end{tabular}
        %\begin{tablenotes}
             All values marked by asterisks indicate the weighted mean periods calculated from multiplets.
        %\end{tablenotes}
    \end{threeparttable}
\end{table*}
%\end{landscape}
%%%%%%%%%%%%%%%%%%%%%%%%%%%%%%%%%%%%

%%%%%%%%%%%%%%%%%%%%%%%%%%%%%%%%%%%%%%%%%%%%%%%%%%%%%%%%%%%%%%%%%%%%%%%%%%%%%%%%%%%%%%%%%%%%%%%%%%%%%%%%%%%%
%\begin{landscape}
\begin{table*}[!ht]
\caption{Same as in Table~\ref{AR2378}, but for AR 12435 and AR 12437. }
\label{AR2435}
 \centering
     \begin{threeparttable}
\begin{tabular}{cccccccc}
 \hline \hline
\multicolumn{4}{c}{AR 12435 } & \multicolumn{4}{c}{AR 12437 } \\
 \hline
M1 & M2 & Averaged periods & M3 & M1 & M2 & Averaged periods & M3  \\
 \hline
\multicolumn{4}{c}{$D_{B_{LOS}}$} & \multicolumn{4}{c}{$D_{B_{LOS}}$} \\
  \hline
-- & \color{cyan}{$16.1\pm0.63$} & \color{cyan}{$16.1\pm0.63$} & -- & \color{cyan}{$17.1\pm0.71$} & \color{cyan}{$17.4\pm0.74$} & \color{cyan}{$17.3\pm0.73$} & -- \\
\color{magenta}{$8.90\pm0.19$} & \color{magenta}{$9.00\pm0.20$}  & \color{magenta}{$8.95\pm0.20$} & \color{magenta}{$8.53\pm0.71$} & -- & \color{magenta}{$7.95\pm0.15$} & \color{magenta}{$7.95\pm0.15$} & \color{magenta}{$7.88\pm0.61$} \\
\color{red}{$6.40\pm0.10$} & \color{red}{$6.45\pm0.10$} & \color{red}{$6.43\pm0.10$} & \color{red}{$6.02\pm0.35$} & -- & -- & -- & -- \\
 \hline
\multicolumn{4}{c}{$D_{B_{r}}$} & \multicolumn{4}{c}{$D_{B_{r}}$} \\
  \hline
-- & \color{magenta}{$7.95\pm0.15$} & \color{magenta}{$7.95\pm0.15$} & -- & -- & -- & -- & -- \\
-- & \color{red}{$5.89\pm0.08$} & \color{red}{$5.89\pm0.08$} & \color{red}{$6.02\pm0.35$} & \color{red}{$5.85\pm0.08$} & \color{red}{$5.89\pm0.08$} & \color{red}{$5.87\pm0.08$} & \color{red}{$6.02\pm0.36$} \\
\color{green}{$4.71\pm0.06$} & \color{green}{$4.77\pm0.06$}\tnote{*} & \color{green}{$4.74\pm0.06$} & \color{green}{$4.88\pm0.23$} & -- & -- & -- & \color{green}{$4.65\pm0.21$} \\
 \hline
\multicolumn{4}{c}{$D_{B_{\theta}}$} & \multicolumn{4}{c}{$D_{B_{\theta}}$} \\
  \hline
-- & \color{magenta}{$7.79\pm0.15$}\tnote{*} & \color{magenta}{$7.79\pm0.15$} & \color{magenta}{$7.88\pm0.61$} & -- & \color{magenta}{$7.95\pm0.15$} & \color{magenta}{$7.95\pm0.15$} & \color{magenta}{$7.88\pm0.61$} \\
\color{red}{$5.95\pm0.09$}\tnote{*} & \color{red}{$6.01\pm0.09$}\tnote{*} & \color{red}{$5.98\pm0.09$} & \color{red}{$6.13\pm0.37$}\tnote{*} & \color{red}{$6.02\pm0.09$} & \color{red}{$6.07\pm0.09$} & \color{red}{$6.04\pm0.09$} & \color{red}{$6.02\pm0.36$} \\
$5.00\pm0.06$ & $5.03\pm0.06$ & $5.02\pm0.06$ & -- & -- & -- & -- & -- \\
\color{green}{$4.75\pm0.05$}\tnote{*} & \color{green}{$4.77\pm0.06$}\tnote{*} & \color{green}{$4.76\pm0.06$} & \color{green}{$4.88\pm0.23$} & \color{green}{$4.82\pm0.06$} & \color{green}{$4.79\pm0.06$} & \color{green}{$4.81\pm0.06$} & \color{green}{$4.88\pm0.23$} \\
-- & $3.40\pm0.06$ & $3.40\pm0.06$ & $3.41\pm0.11$ & -- & -- & -- & -- \\
 \hline
\multicolumn{4}{c}{$D_{B_{\varphi}}$} & \multicolumn{4}{c}{$D_{B_{\varphi}}$} \\
  \hline
-- & \color{magenta}{$7.95\pm0.15$} & \color{magenta}{$7.95\pm0.15$} & \color{magenta}{$7.88\pm0.61$} & \color{magenta}{$8.53\pm0.18$} & \color{magenta}{$8.62\pm0.18$} & \color{magenta}{$8.58\pm0.18$} & --  \\
\color{red}{$5.85\pm0.08$} & \color{red}{$5.89\pm0.08$} & \color{red}{$5.87\pm0.08$} & \color{red}{$5.69\pm0.32$} & \color{red}{$5.89\pm0.09$}\tnote{*} & \color{red}{$5.96\pm0.09$}\tnote{*} & \color{red}{$5.93\pm0.09$} & \color{red}{$5.86\pm0.34$}\tnote{*} \\
\color{green}{$4.82\pm0.06$} & \color{green}{$4.79\pm0.06$} & \color{green}{$4.81\pm0.06$} & \color{green}{$4.65\pm0.21$} & -- & \color{green}{$4.68\pm0.05$} & \color{green}{$4.68\pm0.05$} & \color{green}{$4.65\pm0.21$} \\
  \hline
\end{tabular}
        %\begin{tablenotes}
             All values marked by asterisks indicate the weighted mean periods calculated from multiplets.
        %\end{tablenotes}
    \end{threeparttable}
\end{table*}
%\end{landscape}
%%%%%%%%%%%%%%%%%%%%%%%%%%%%%%%%%%%%%%

%%%%%%%%%%%%%%%%%%%%%%%%%%%%%%%%%%%%%%%%%%%%%%%%%%%%%%%%%%%%%%%%%%%%%%%%%%%%%%%%%%%%%%%%%%%%%%%%%%%%%%%%%%%%%
\begin{table}[!ht]
\caption{Same as in Table~\ref{AR2378}, but for AR 12524.}
\label{AR2524}
 \centering
     \begin{threeparttable}
\begin{tabular}{cccc}
 \hline \hline
M1 & M2 & Averaged periods & M3  \\
 \hline
\multicolumn{4}{c}{$D_{B_{LOS}}$} \\
  \hline
-- & \color{blue}{$3.73\pm0.04$}\tnote{*}   & \color{blue}{$3.73\pm0.04$}  & -- \\
-- & $2.84\pm0.04$                                     & $2.84\pm0.04$                       & $2.84\pm0.08$ \\
 \hline
\multicolumn{4}{c}{$D_{B_{r}}$} \\
  \hline
\color{red}{$6.02\pm0.09$}   & \color{red}{$6.06\pm0.09$}      & \color{red}{$6.04\pm0.09$}   & --   \\
\color{green}{$4.60\pm0.05$} & \color{green}{$4.58\pm0.05$} & \color{green}{$4.59\pm0.05$} & \color{green}{$4.65\pm0.21$}  \\
$3.98\pm0.04$                    & $3.96\pm0.04$                         & $3.97\pm0.04$                & --   \\
 \hline
\multicolumn{4}{c}{$D_{B_{\theta}}$} \\
  \hline
-- & \color{red}{$5.73\pm0.08$}    & \color{red}{$5.73\pm0.08$}     & \color{red}{$5.69\pm0.32$}  \\
-- & \color{green}{$4.58\pm0.06$} & \color{green}{$4.58\pm0.06$} & \color{green}{$4.88\pm0.23$} \\
 \hline
\multicolumn{4}{c}{$D_{B_{\varphi}}$} \\
  \hline
-- & -- & -- & -- \\
  \hline
\end{tabular}
        %\begin{tablenotes}
             All values marked by asterisks indicate the weighted mean periods calculated from multiplets.
        %\end{tablenotes}
    \end{threeparttable}
\end{table}

\begin{table}[h!]
\caption{Detected periods (with corresponding errors) for the radial unsigned magnetic flux for each active region. The formatting is the same as in previous tables. }
\label{flux}
 \centering
      \begin{threeparttable}
\begin{tabular}{cccc}
 \hline \hline
M1 & M2 & Averaged periods & M3  \\
 \hline
\multicolumn{4}{c}{AR 12378} \\
 \hline
\color{red}{$6.21\pm0.09$} & \color{red}{$6.25\pm0.09$} & \color{red}{$6.23\pm0.09$} & \color{red}{$6.02\pm0.36$} \\
 \hline
\multicolumn{4}{c}{AR 12381} \\
 \hline
-- & -- & -- & -- \\
 \hline
\multicolumn{4}{c}{AR 12435} \\
 \hline
-- & $7.38\pm0.14$  & $7.38\pm0.14$                & $7.31\pm0.52$ \\
 \hline
\multicolumn{4}{c}{AR 12437} \\
 \hline
\color{red}{$6.03\pm0.09$}\tnote{*}  & -- & \color{red}{$6.03\pm0.09$}   & \color{red}{$6.02\pm0.36$} \\
\color{green}{$4.80\pm0.06$} & \color{green}{$4.91\pm0.06$}  & \color{green}{$4.86\pm0.06$} & \color{green}{$4.65\pm0.21$} \\
 \hline
\multicolumn{4}{c}{AR 12524} \\
 \hline
-- & \color{red}{$6.07\pm0.09$}   & \color{red}{$6.07\pm0.09$}   & -- \\
-- & \color{green}{$4.58\pm0.05$} & \color{green}{$4.58\pm0.05$} & \color{green}{$4.65\pm0.21$} \\
-- & $3.96\pm0.04$                & $3.96\pm0.04$                & -- \\
 \hline
\end{tabular}
        %\begin{tablenotes}
             All values marked by asterisks indicate the weighted mean periods calculated from multiplets.
        %\end{tablenotes}
    \end{threeparttable}
\end{table}
%%%%%%%%%%%%%%%%%%%%%%%%%%%%%%%%%%%%%
%%%%%%%%%%%%%%%%%%%%%%%%%%%%%%%%%%%%%%%%%%%%%%%%%%%%%%%%%%%%%%%%%%%%%%%%%%%%%%%%%%%%%%%%%%%%%%%%%%%%%%%%%%%%

%\newpage
%%%%%%%%%%%%%%%%%%%%%%%%%%%%%%%%%%%%%%%%%%%%%%%%%%%%%%%%%%%
\section{Example eigenspectra}\label{appfigures}
\begin{landscape}
\begin{figure}[!ht]
\begin{minipage}[h]{0.31\linewidth}\centering
{\includegraphics[width=0.89\textwidth]{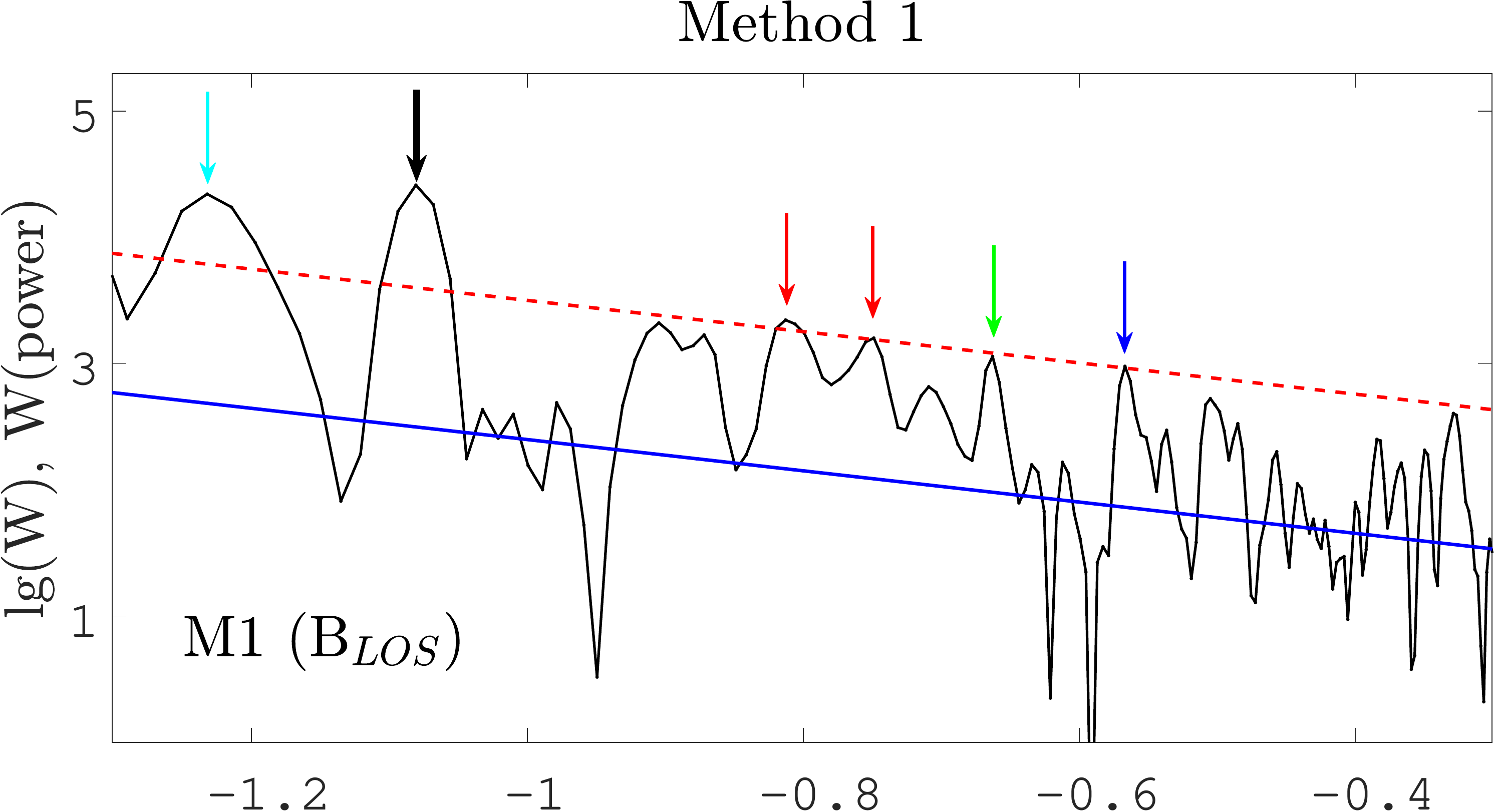}}
{\includegraphics[width=0.89\textwidth]{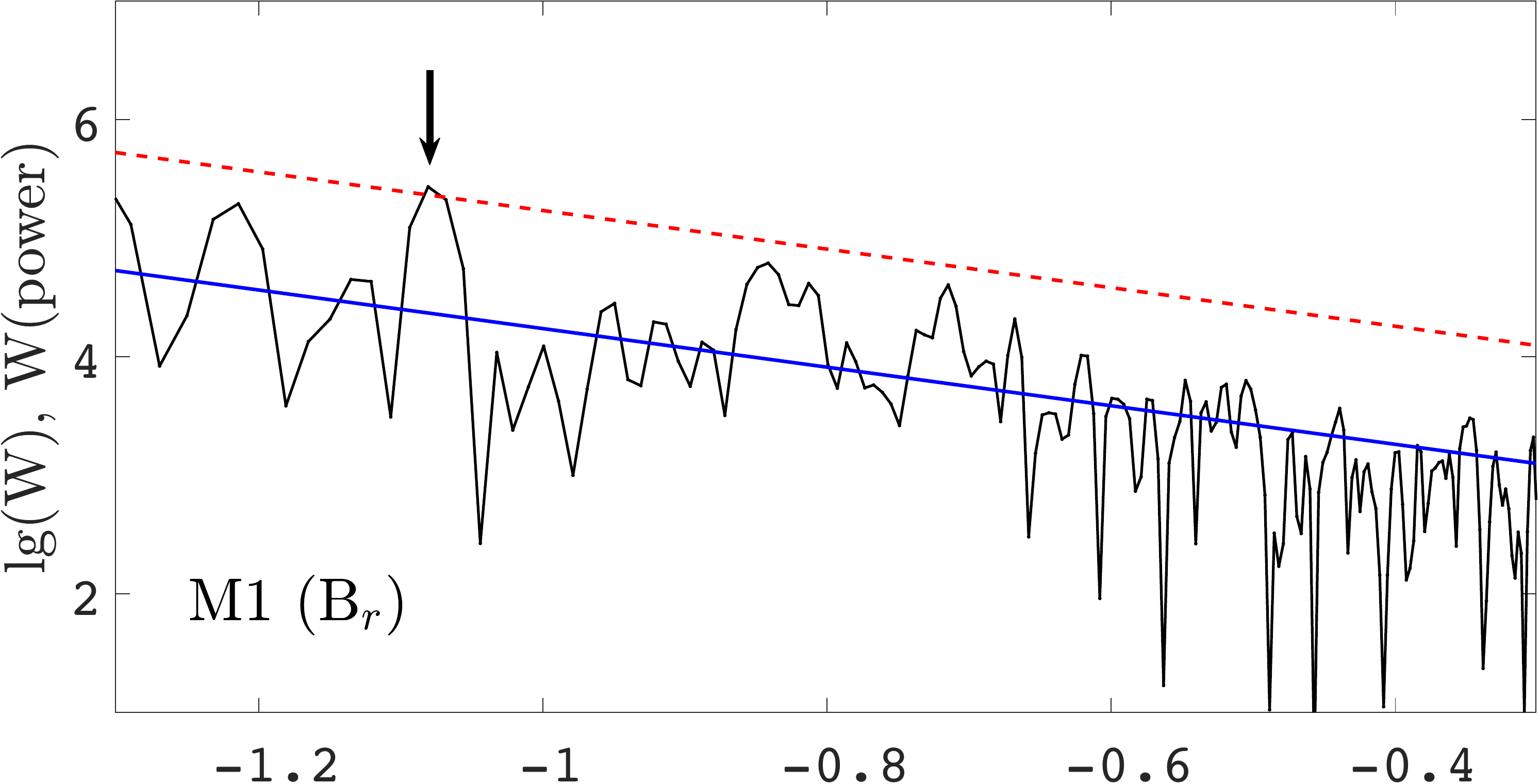}}
{\includegraphics[width=0.89\textwidth]{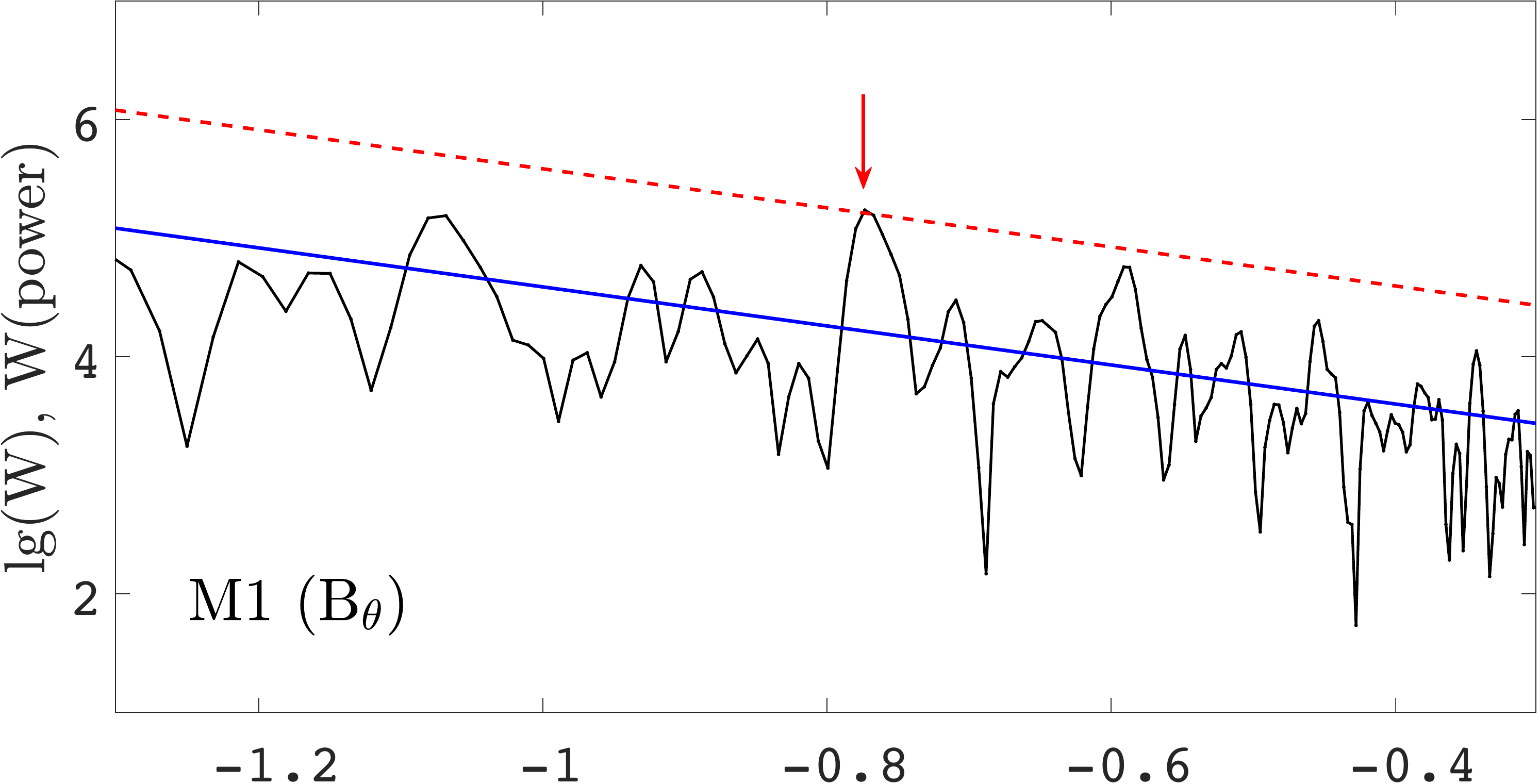}}
{\includegraphics[width=0.89\textwidth]{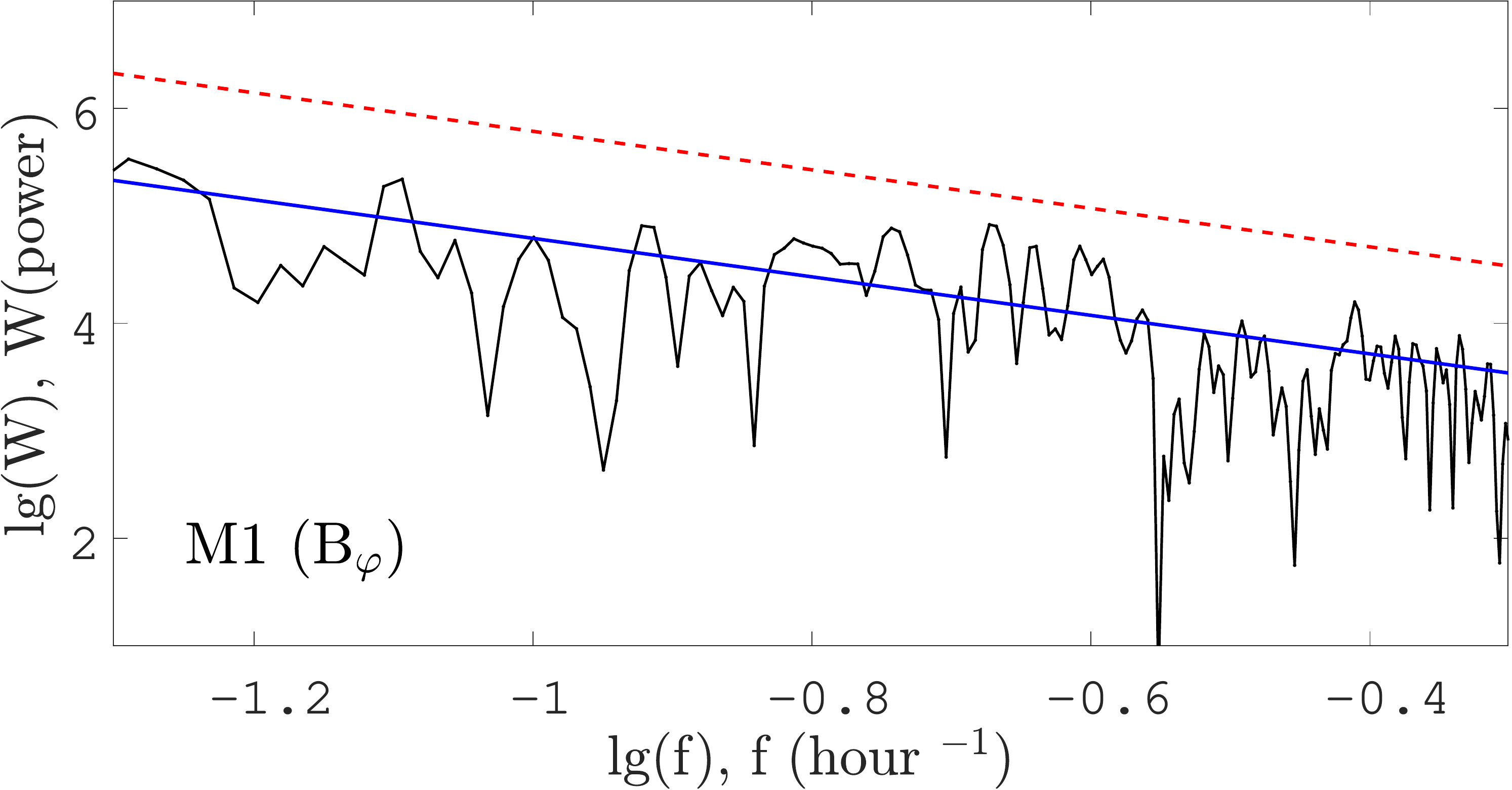}}
\end{minipage}
\begin{minipage}[h]{0.31\linewidth}\centering
{\includegraphics[width=0.86\textwidth]{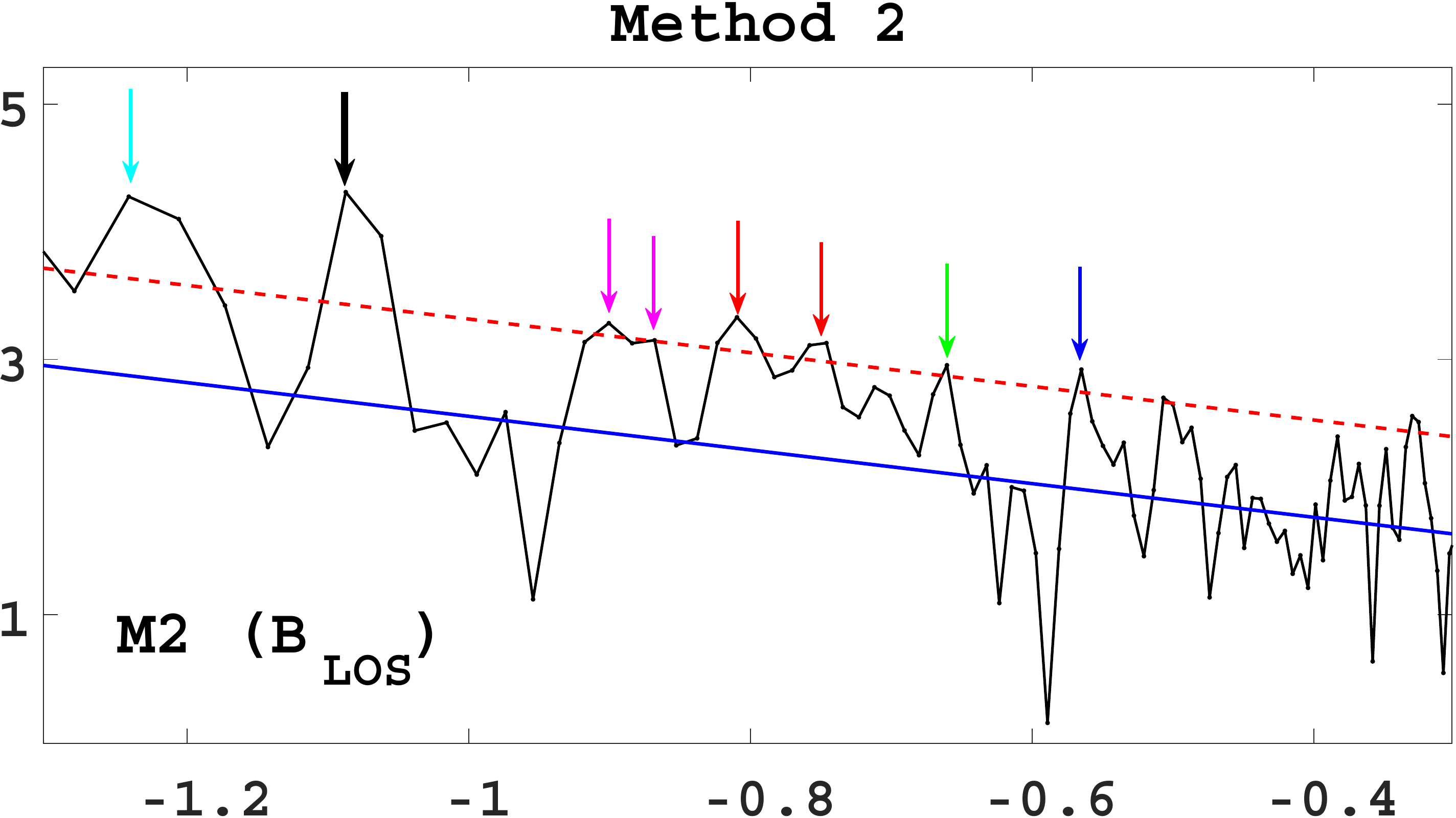}}
{\includegraphics[width=0.86\textwidth]{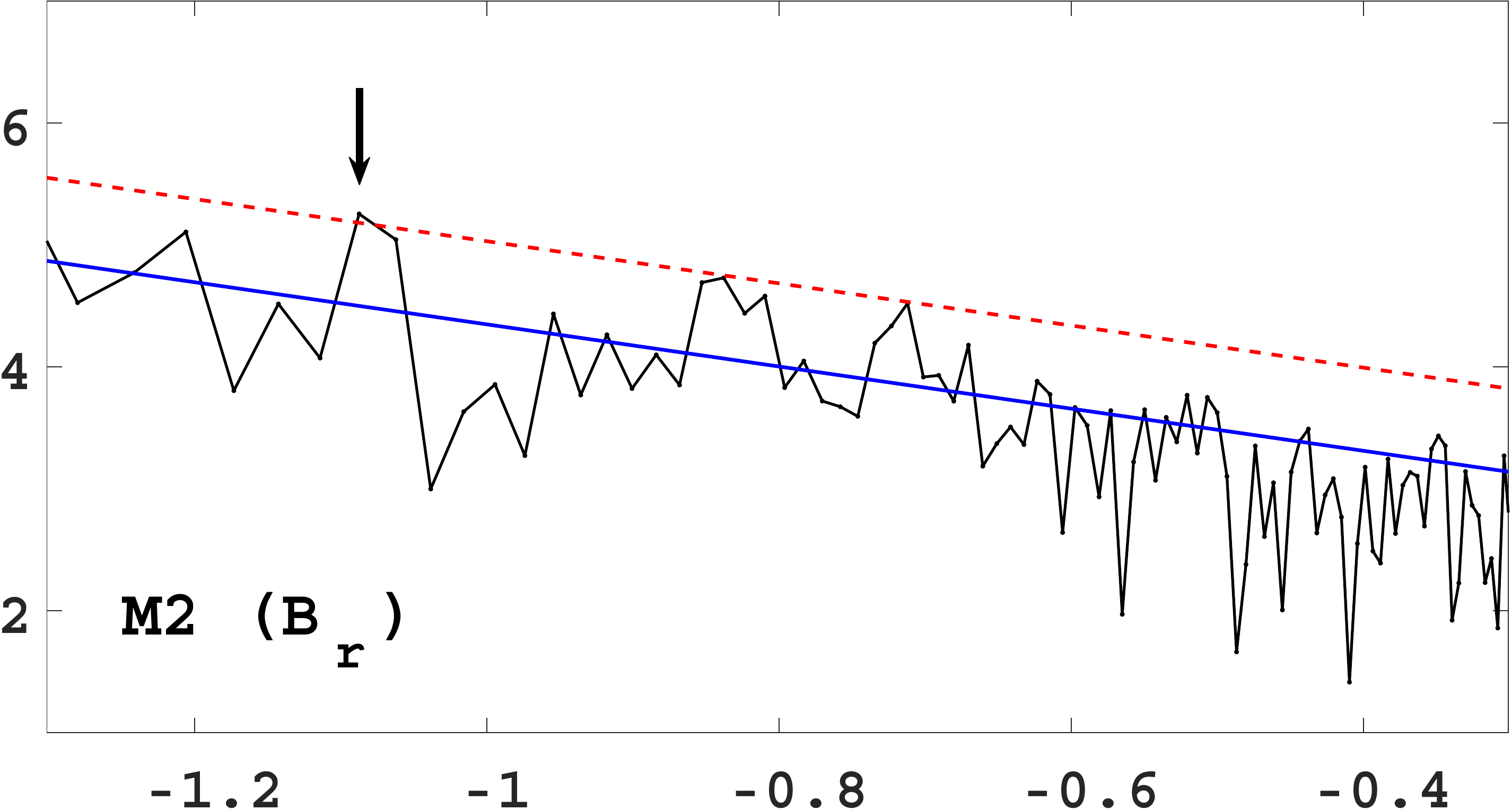}}
{\includegraphics[width=0.86\textwidth]{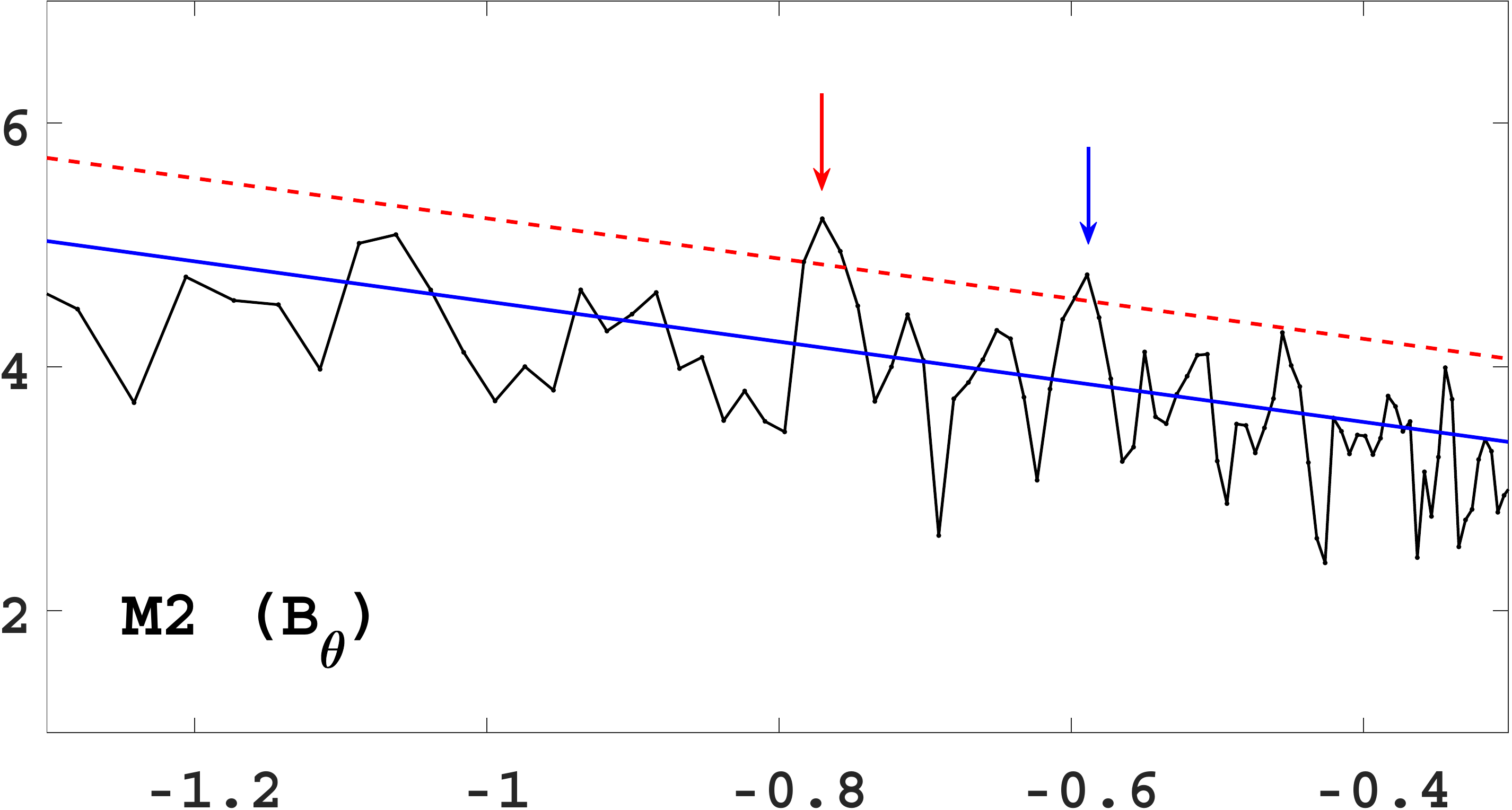}}
{\includegraphics[width=0.86\textwidth]{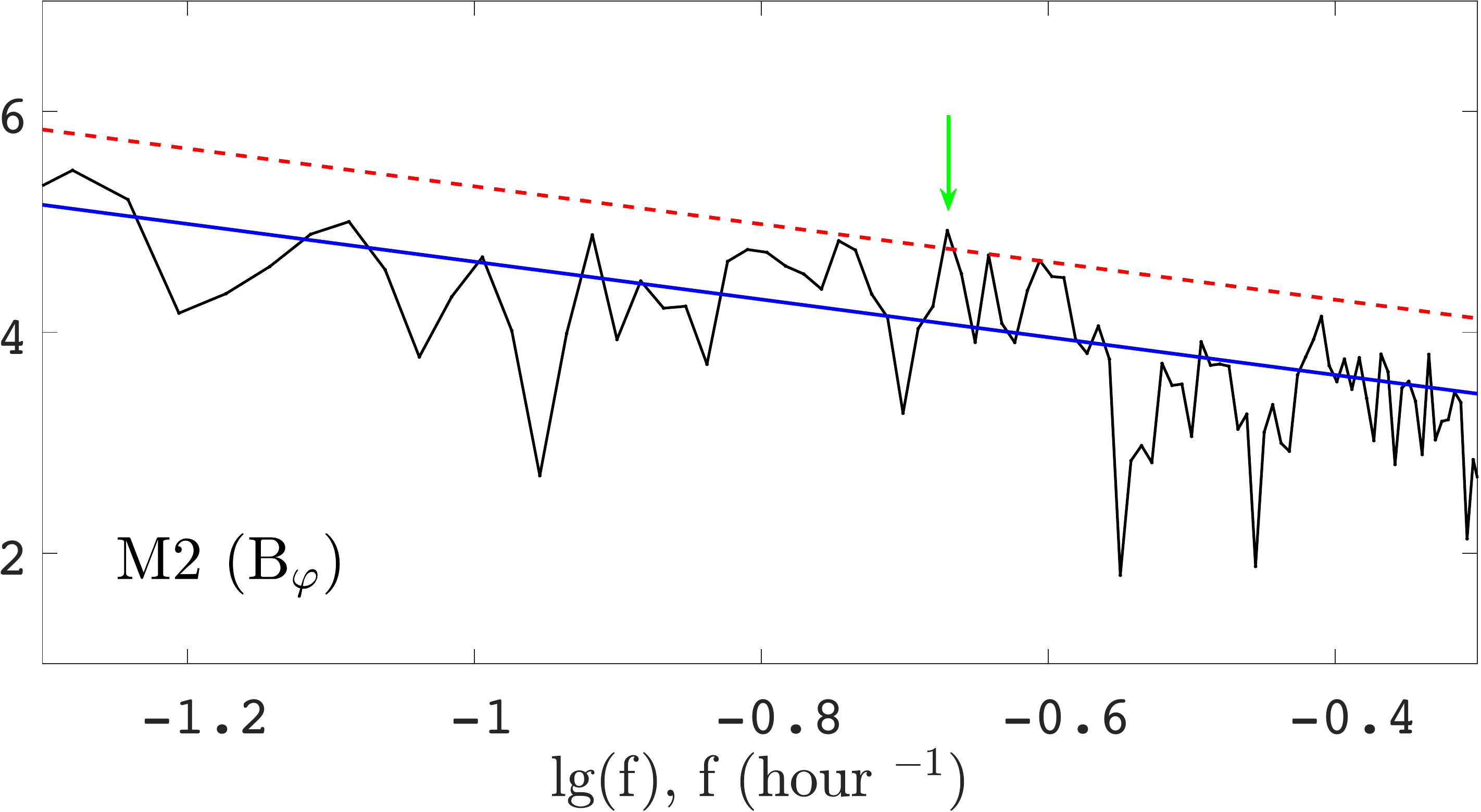}}
\end{minipage}
\begin{minipage}[h]{0.31\linewidth}\centering
{\includegraphics[width=0.87\textwidth]{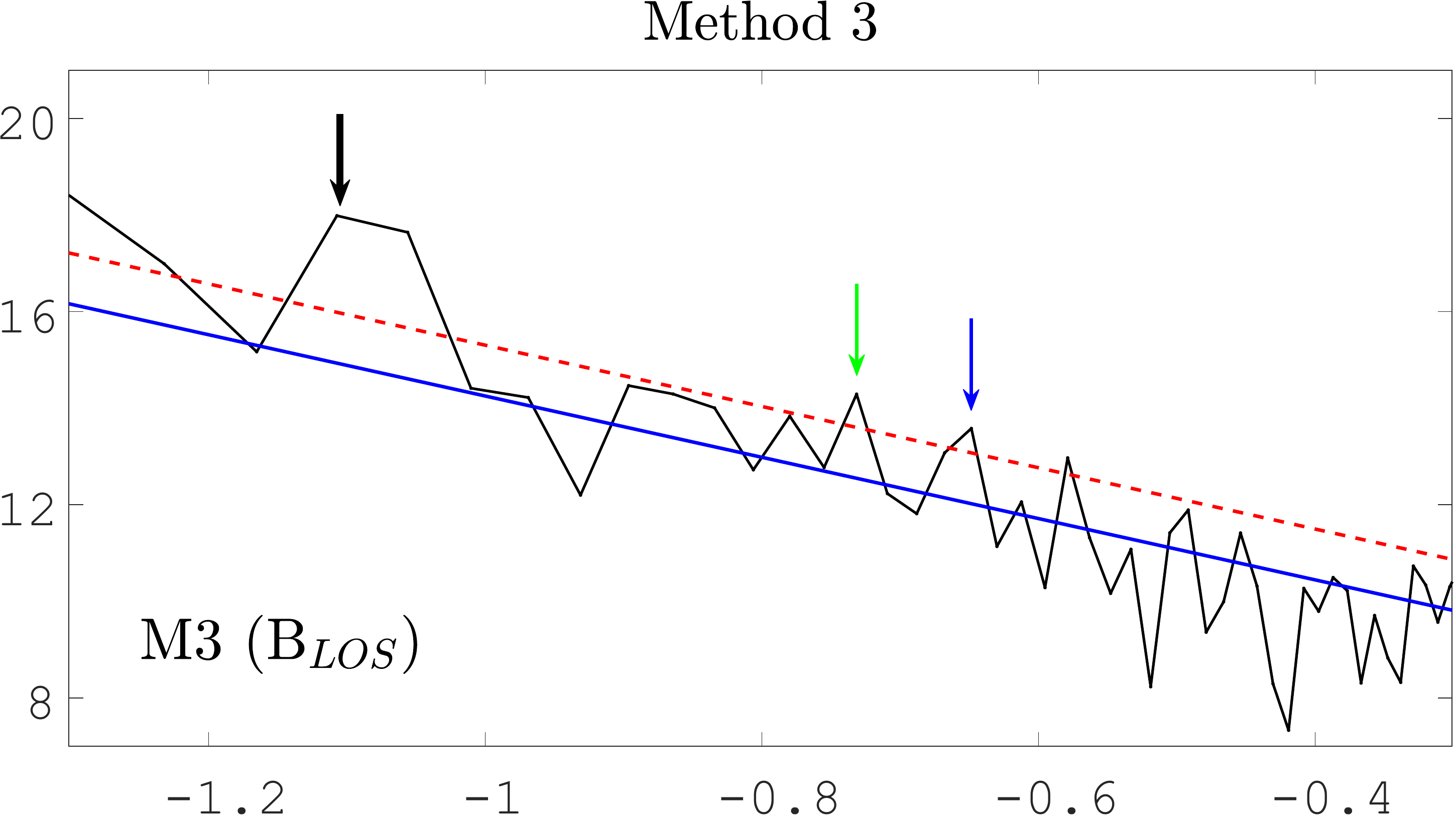}}
{\includegraphics[width=0.87\textwidth]{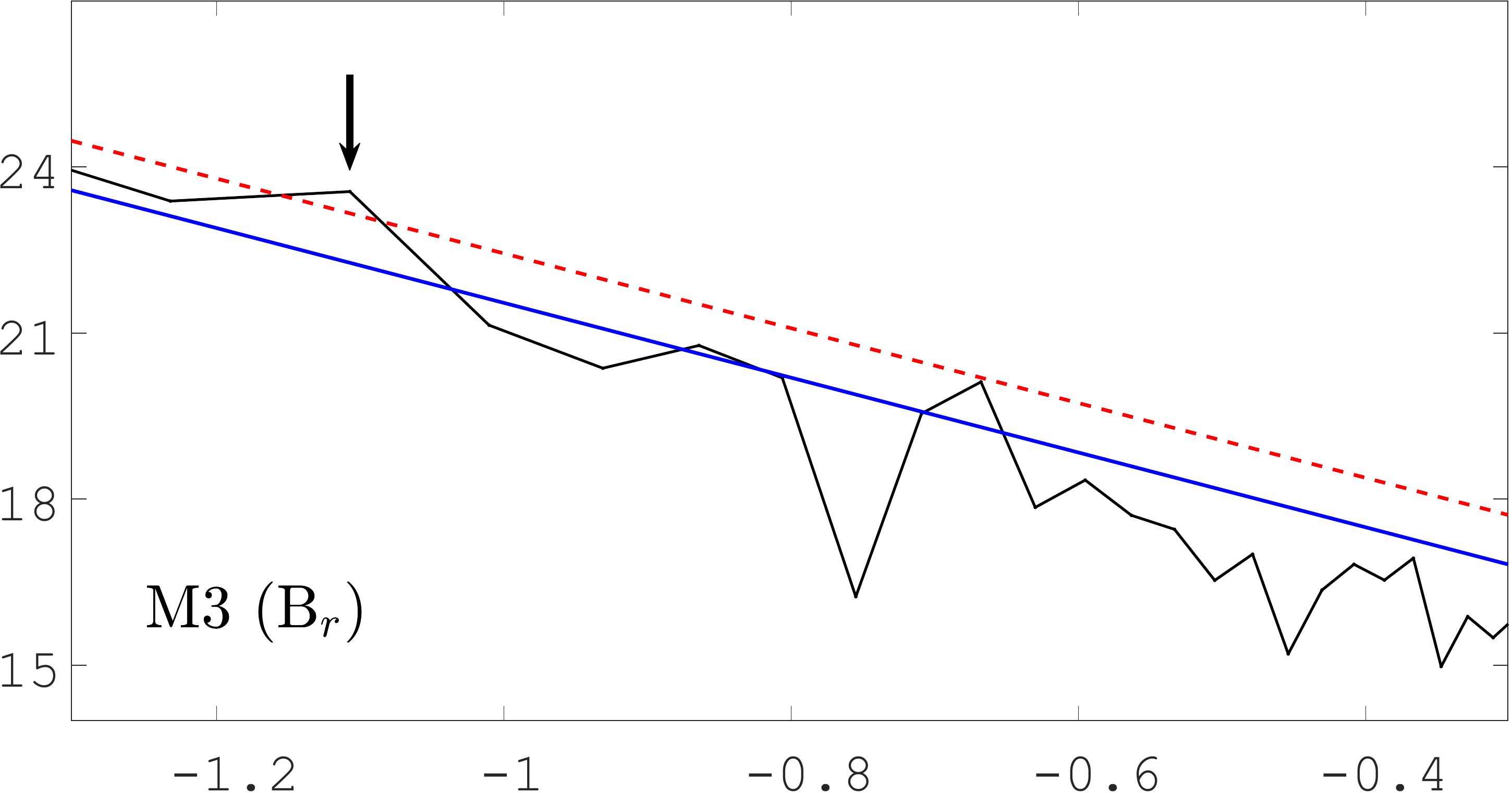}}
{\includegraphics[width=0.87\textwidth]{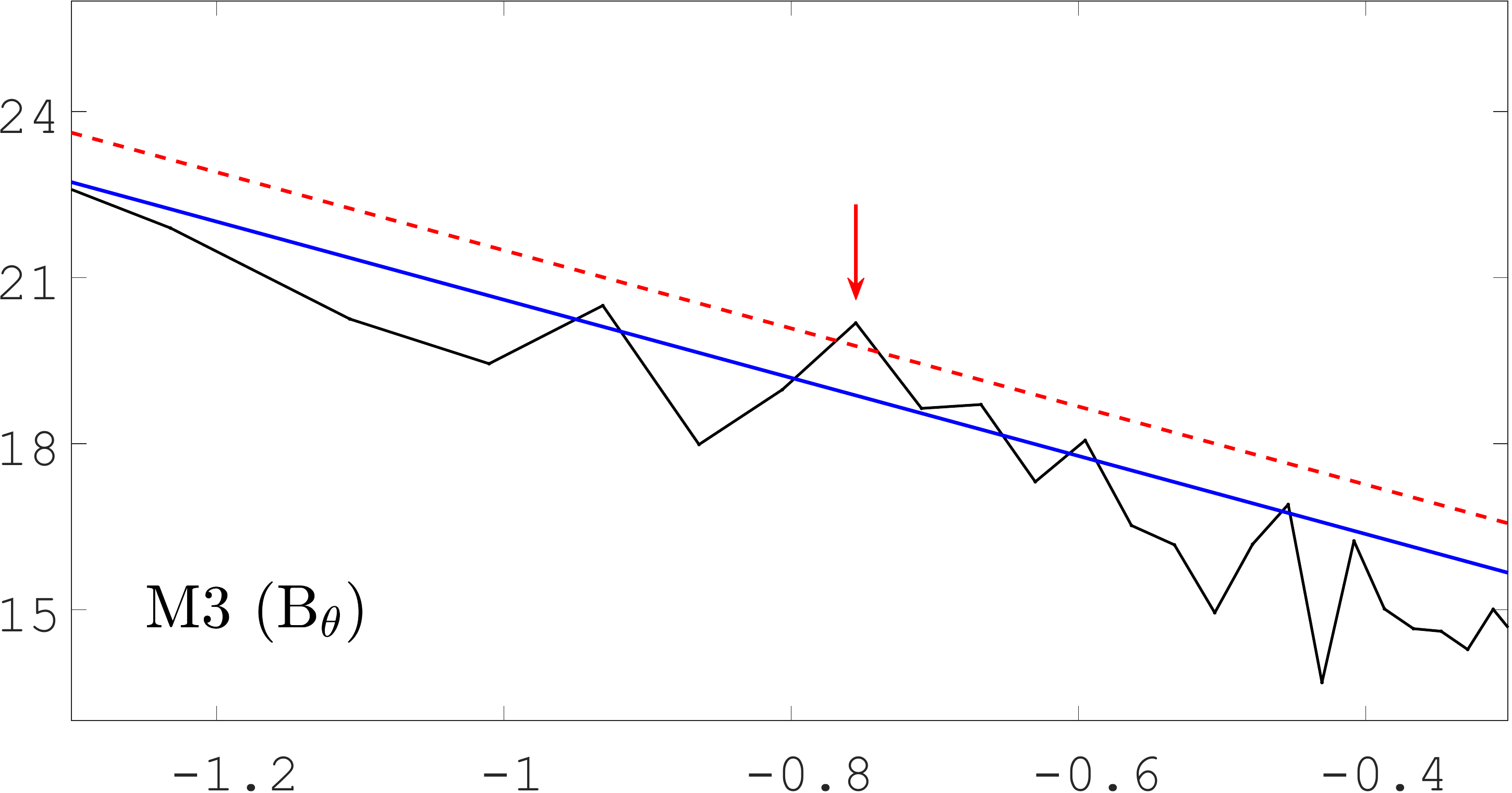}}
{\includegraphics[width=0.87\textwidth]{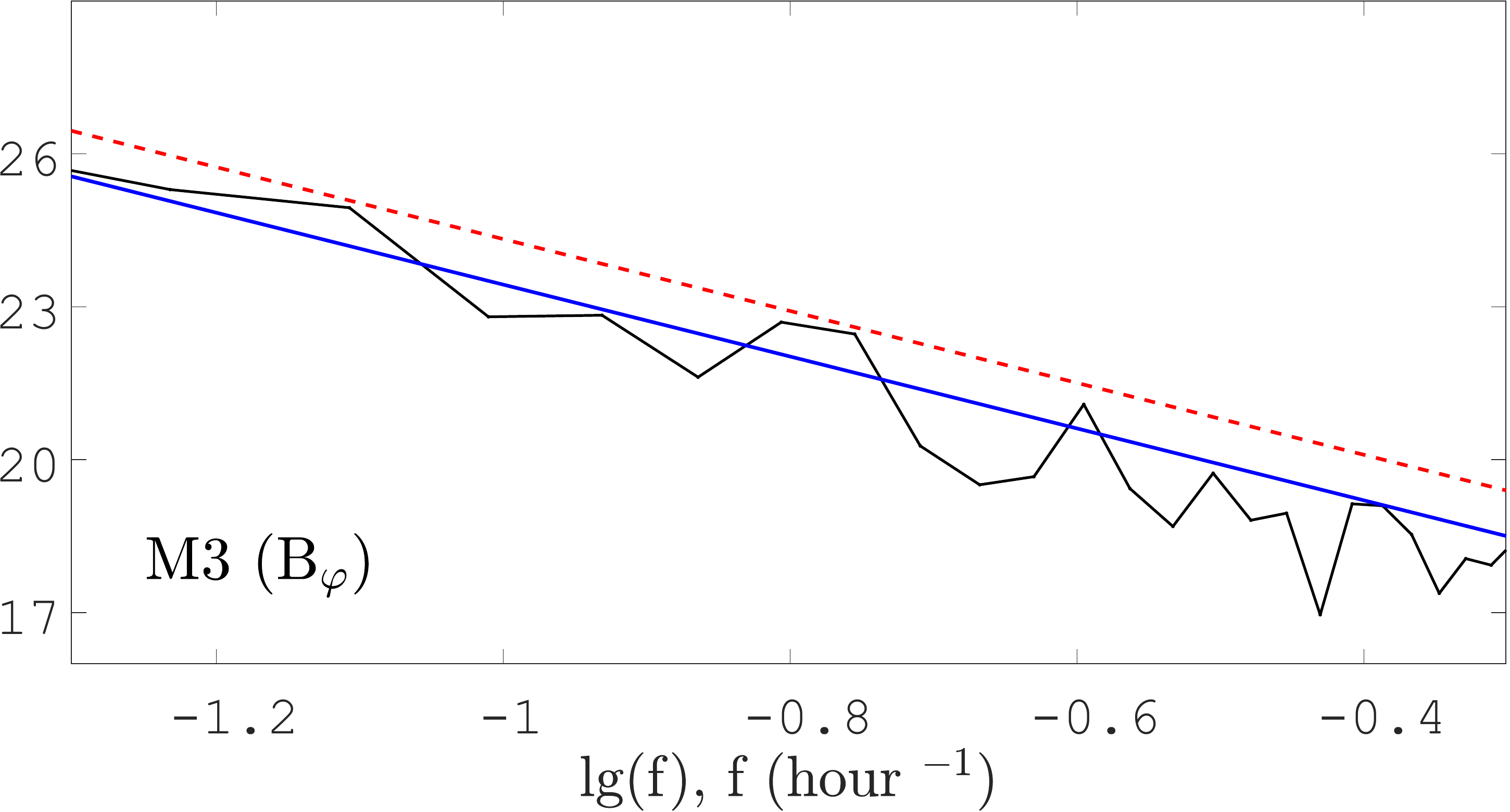}}
\end{minipage}
\caption{Example of power spectra obtained for the area of AR 12381 using methods M1, M2, and M3  for $B_{LOS}$ (the first row of panels from top), $B_r$ (the second row), $B_{\theta}$ (the third row), $B_{\varphi}$ (the fourth row) data. The  solid blue line presents the model line ($\lg W_{model}=m\lg f+b$) and the red dashed line indicates the 95\% confidence level. The arrows indicate the significant peaks for each case, accordingly. The colouring of all arrows follows that used in  Table~\ref{AR2378}. The thick black arrows denote the artefact periods of 12 hours \citep{Liu12}. }
\label{M1}
\end{figure}
\end{landscape}

%%%%%%%%%%%%%%%%%%%%%%%%%%%%%%%%%%%%%%%%%%%%%%%%%%%%%%%%%%%
Here, we provide an example of the eigenspectra of the AR 12381 area.
Fig.~\ref{M1} shows the obtained frequency spectra for AR 12381 with all three methods that are shown with the detected peaks, as well as the 12-hour artefact frequencies \citep{Liu12}.

%%%%%%%------------------------------------------------------------M1, M2, M3 and M4 of AR 12381
%%%------------------------

%%%%%%%%%%%%%%%%%%%%%%%%%%%%%%%%%%%%%%%%%%
\bibliography{mybib}

\begin{thebibliography}{45}
\expandafter\ifx\csname natexlab\endcsname\relax\def\natexlab#1{#1}\fi

\bibitem[{{Abramov-Maximov} {et~al.}(2013){Abramov-Maximov}, {Efremov},
  {Parfinenko}, {Solov'ev}, \& {Shibasaki}}]{Abramov2013}
{Abramov-Maximov}, V.~E., {Efremov}, V.~I., {Parfinenko}, L.~D., {Solov'ev},
  A.~A., \& {Shibasaki}, K. 2013, \pasj, 65, S12

\bibitem[{{Andries} {et~al.}(2009){Andries}, {van Doorsselaere}, {Roberts},
  {Verth}, {Verwichte}, \& {Erd{\'e}lyi}}]{2009SSRv..149....3A}
{Andries}, J., {van Doorsselaere}, T., {Roberts}, B., {et~al.} 2009, \ssr, 149,
  3

\bibitem[{{Appourchaux}(2003)}]{Appourchaux2003}
{Appourchaux}, T. 2003, \aap, 412, 903

\bibitem[{{Baiesi} {et~al.}(2008){Baiesi}, {Maes}, \&
  {Shergelashvili}}]{Baiesi2008}
{Baiesi}, M., {Maes}, C., \& {Shergelashvili}, B.~M. 2008, Physica A
  Statistical Mechanics and its Applications, 387, 167

\bibitem[{Belkasim {et~al.}(1991)Belkasim, Shridhar, \& Ahmadi}]{BELKASIM1991}
Belkasim, S., Shridhar, M., \& Ahmadi, M. 1991, Pattern Recognition, 24, 1117

\bibitem[{{Bobra} {et~al.}(2014){Bobra}, {Sun}, {Hoeksema}, {Turmon}, {Liu},
  {Hayashi}, {Barnes}, \& {Leka}}]{Bobra14}
{Bobra}, M.~G., {Sun}, X., {Hoeksema}, J.~T., {et~al.} 2014, \solphys, 289,
  3549

\bibitem[{Bogart {et~al.}(2011{\natexlab{a}})Bogart, Baldner, Basu, Haber, \&
  Rabello-Soares}]{Bogart2011a}
Bogart, R.~S., Baldner, C., Basu, S., Haber, D.~A., \& Rabello-Soares, M.~C.
  2011{\natexlab{a}}, Journal of Physics: Conference Series, 271, 012008

\bibitem[{Bogart {et~al.}(2011{\natexlab{b}})Bogart, Baldner, Basu, Haber, \&
  Rabello-Soares}]{Bogart2011b}
Bogart, R.~S., Baldner, C., Basu, S., Haber, D.~A., \& Rabello-Soares, M.~C.
  2011{\natexlab{b}}, Journal of Physics: Conference Series, 271, 012009

\bibitem[{{Centeno} {et~al.}(2006){Centeno}, {Collados}, \& {Trujillo
  Bueno}}]{centeno06}
{Centeno}, R., {Collados}, M., \& {Trujillo Bueno}, J. 2006, \apj, 640, 1153

\bibitem[{{Chorley} {et~al.}(2010){Chorley}, {Hnat}, {Nakariakov}, {Inglis}, \&
  {Bakunina}}]{Chorley10}
{Chorley}, N., {Hnat}, B., {Nakariakov}, V.~M., {Inglis}, A.~R., \& {Bakunina},
  I.~A. 2010, \aap, 513, A27

\bibitem[{{Dumbadze} {et~al.}(2017){Dumbadze}, {Shergelashvili}, {Kukhianidze},
  {Ramishvili}, {Zaqarashvili}, {Khodachenko}, {Gurgenashvili}, {Poedts}, \&
  {De Causmaecker}}]{Dumbadze17}
{Dumbadze}, G., {Shergelashvili}, B.~M., {Kukhianidze}, V., {et~al.} 2017,
  \aap, 597, A93

\bibitem[{{Efremov} {et~al.}(2007){Efremov}, {Parfinenko}, \&
  {Solov'ev}}]{efremov07}
{Efremov}, V.~I., {Parfinenko}, L.~D., \& {Solov'ev}, A.~A. 2007, Astronomy
  Reports, 51, 401

\bibitem[{{Fleck} \& {Schmitz}(1991)}]{fleckschmits91}
{Fleck}, B. \& {Schmitz}, F. 1991, \aap, 250, 235

\bibitem[{Flusser(2007)}]{flusser07}
Flusser, J. 2007, International Journal of Computer, Electrical, Automation,
  Control and Information Engineering, 1, 3708

\bibitem[{{Flusser} \& {Suk}(1994)}]{Flusser1994}
{Flusser}, J. \& {Suk}, T. 1994, IEEE Transactions on Geoscience and Remote
  Sensing, 32, 382

\bibitem[{{Gelfreikh} {et~al.}(2006){Gelfreikh}, {Nagovitsyn}, \&
  {Nagovitsyna}}]{gelfreichetal06}
{Gelfreikh}, G.~B., {Nagovitsyn}, Y.~A., \& {Nagovitsyna}, E.~Y. 2006, \pasj,
  58, 29

\bibitem[{{Goldvarg} {et~al.}(2005){Goldvarg}, {Nagovitsyn}, \&
  {Solov'Ev}}]{goldvarg05}
{Goldvarg}, T.~B., {Nagovitsyn}, Y.~A., \& {Solov'Ev}, A.~A. 2005, Astronomy
  Letters, 31, 414

\bibitem[{{Gopasyuk}(2004)}]{gopasyuk04}
{Gopasyuk}, O.~S. 2004, in IAU Symposium, Vol. 223, Multi-Wavelength
  Investigations of Solar Activity, ed. A.~V. {Stepanov}, E.~E.
  {Benevolenskaya}, \& A.~G. {Kosovichev}, 249--250

\bibitem[{Goshtasby(1985)}]{Goshtasby1985}
Goshtasby, A. 1985, IEEE Transactions on Geoscience and Remote Sensing, 7, 338

\bibitem[{{Hoeksema} {et~al.}(2014){Hoeksema}, {Liu}, {Hayashi}, {Sun},
  {Schou}, {Couvidat}, {Norton}, {Bobra}, {Centeno}, {Leka}, {Barnes}, \&
  {Turmon}}]{Hoeksema14}
{Hoeksema}, J.~T., {Liu}, Y., {Hayashi}, K., {et~al.} 2014, \solphys, 289, 3483

\bibitem[{Hu(1962)}]{hu62}
Hu, M.-K. 1962, {IRE} Transactions on Information Theory, 8

\bibitem[{{Jaeggli} \& {Norton}(2016)}]{Jaeggli2016}
{Jaeggli}, S.~A. \& {Norton}, A.~A. 2016, \apjl, 820, L11

\bibitem[{{Khutsishvili} {et~al.}(1998){Khutsishvili}, {Kvernadze}, \&
  {Sikharulidze}}]{khutsishvili98}
{Khutsishvili}, E., {Kvernadze}, T., \& {Sikharulidze}, M. 1998, \solphys, 178,
  271

\bibitem[{{Kuridze} {et~al.}(2009){Kuridze}, {Zaqarashvili}, {Shergelashvili},
  \& {Poedts}}]{kuridze09}
{Kuridze}, D., {Zaqarashvili}, T.~V., {Shergelashvili}, B.~M., \& {Poedts}, S.
  2009, \aap, 505, 763

\bibitem[{{Liu} {et~al.}(2012){Liu}, {Hoeksema}, {Scherrer}, {Schou},
  {Couvidat}, {Bush}, {Duvall}, {Hayashi}, {Sun}, \& {Zhao}}]{Liu12}
{Liu}, Y., {Hoeksema}, J.~T., {Scherrer}, P.~H., {et~al.} 2012, \solphys, 279,
  295

\bibitem[{{Maes} {et~al.}(2009){Maes}, {Neto{\v{c}}n{\'y}}, \&
  {Shergelashvili}}]{Maes2009}
{Maes}, C., {Neto{\v{c}}n{\'y}}, K., \& {Shergelashvili}, B.~M. 2009, \pre, 80,
  011121

\bibitem[{{Moradi}(2012)}]{Moradi12}
{Moradi}, H. 2012, Astronomische Nachrichten, 333, 1003

\bibitem[{{Nagovitsyn} \& {Nagovitsyna}(2011)}]{Nagovitsyna2011}
{Nagovitsyn}, Y.~A. \& {Nagovitsyna}, E.~Y. 2011, Geomagnetism and Aeronomy,
  51, 1049

\bibitem[{{Parker}(1979)}]{parker1979}
{Parker}, E.~N. 1979, \apj, 234, 333

\bibitem[{Press {et~al.}(2007)Press, Teukolsky, Vetterling, \&
  Flannery}]{press2007}
Press, W., Teukolsky, S., Vetterling, W., \& Flannery, B. 2007, Numerical
  Recipes 3rd Edition: The Art of Scientific Computing (Cambridge University
  Press)

\bibitem[{Prokop \& Reeves(1992)}]{Prokop92}
Prokop, R.~J. \& Reeves, A.~P. 1992, CVGIP: Graph. Models Image Process., 54,
  438

\bibitem[{{Pugh} {et~al.}(2017){Pugh}, {Broomhall}, \& {Nakariakov}}]{Pugh2017}
{Pugh}, C.~E., {Broomhall}, A.~M., \& {Nakariakov}, V.~M. 2017, \aap, 602, A47

\bibitem[{{Rincon} \& {Rieutord}(2018)}]{Rincon2018}
{Rincon}, F. \& {Rieutord}, M. 2018, Living Reviews in Solar Physics, 15, 6

\bibitem[{{Scherrer} {et~al.}(2012){Scherrer}, {Schou}, {Bush}, {Kosovichev},
  {Bogart}, {Hoeksema}, {Liu}, {Duvall}, {Zhao}, {Title}, {Schrijver},
  {Tarbell}, \& {Tomczyk}}]{Scherrer12}
{Scherrer}, P.~H., {Schou}, J., {Bush}, R.~I., {et~al.} 2012, \solphys, 275,
  207

\bibitem[{{Schou} {et~al.}(2012){Schou}, {Scherrer}, {Bush}, {Wachter},
  {Couvidat}, {Rabello-Soares}, {Bogart}, {Hoeksema}, {Liu}, {Duvall}, {Akin},
  {Allard}, {Miles}, {Rairden}, {Shine}, {Tarbell}, {Title}, {Wolfson},
  {Elmore}, {Norton}, \& {Tomczyk}}]{Schou12}
{Schou}, J., {Scherrer}, P.~H., {Bush}, R.~I., {et~al.} 2012, \solphys, 275,
  229

\bibitem[{{Shergelashvili} {et~al.}(2007){Shergelashvili}, {Maes}, {Poedts}, \&
  {Zaqarashvili}}]{Shergelashvili2007}
{Shergelashvili}, B.~M., {Maes}, C., {Poedts}, S., \& {Zaqarashvili}, T.~V.
  2007, \pre, 76, 046404

\bibitem[{{Shergelashvili} \& {Poedts}(2005)}]{shergelashvili05}
{Shergelashvili}, B.~M. \& {Poedts}, S. 2005, \aap, 438, 1083

\bibitem[{{Shergelashvili} {et~al.}(2005){Shergelashvili}, {Zaqarashvili},
  {Poedts}, \& {Roberts}}]{Shergelashvili2005}
{Shergelashvili}, B.~M., {Zaqarashvili}, T.~V., {Poedts}, S., \& {Roberts}, B.
  2005, \aap, 429, 767

\bibitem[{{Smirnova} {et~al.}(2013){Smirnova}, {Riehokainen}, {Solov'ev},
  {Kallunki}, {Zhiltsov}, \& {Ryzhov}}]{Smirnova13}
{Smirnova}, V., {Riehokainen}, A., {Solov'ev}, A., {et~al.} 2013, \aap, 552,
  A23

\bibitem[{{Solov'ev} \& {Kirichek}(2008)}]{solovev08}
{Solov'ev}, A.~A. \& {Kirichek}, E.~A. 2008, Astrophysical Bulletin, 63, 169

\bibitem[{{Sun}(2013)}]{jsoc2013}
{Sun}, X. 2013, arXiv e-prints, arXiv:1309.2392

\bibitem[{{Terradas} {et~al.}(2008){Terradas}, {Arregui}, {Oliver},
  {Ballester}, {Andries}, \& {Goossens}}]{Terradas2008}
{Terradas}, J., {Arregui}, I., {Oliver}, R., {et~al.} 2008, \apj, 679, 1611

\bibitem[{{Thomas} {et~al.}(1984){Thomas}, {Cram}, \& {Nye}}]{thomas84}
{Thomas}, J.~H., {Cram}, L.~E., \& {Nye}, A.~H. 1984, \apj, 285, 368

\bibitem[{Tsirikolias \& Mertzios(1993)}]{TSIRIKOLIAS1993}
Tsirikolias, K. \& Mertzios, B. 1993, Pattern Recognition, 26, 877

\bibitem[{{Vaughan}(2005)}]{Vaughan2005}
{Vaughan}, S. 2005, \aap, 431, 391

\end{thebibliography}
\end{document}